\documentclass[manuscript]{acmart}
\AtBeginDocument{%
  }

\usepackage{framed}
\usepackage{subcaption}
\usepackage{colortbl}
\definecolor{rowshade}{gray}{0.94}

\newcommand{\baseline}[1]{#1}
\newcommand{\style}[1]{#1}
\newcommand{\context}[1]{#1}
\newcommand{\combined}[1]{#1}

\definecolor{layerConst}{HTML}{3A3F47}
\definecolor{facStyle}{HTML}{2F7D4F}
\definecolor{facContext}{HTML}{6E5FA8}
\definecolor{facInter}{HTML}{5F6570}
\newcommand{\facstyle}[1]{#1}
\newcommand{\faccontext}[1]{#1}
\newcommand{\lyrstyle}{\texttt{<style>}}
\newcommand{\lyrcontext}{\texttt{<context>}}
\newcommand{\lyrrec}{\texttt{<recommendation>}}
\newcommand{\lyrtask}{\texttt{<task>}}

\setcopyright{acmlicensed}
\copyrightyear{2018}
\acmYear{2018}
\acmDOI{XXXXXXX.XXXXXXX}
\acmConference[Conference acronym 'XX]{Make sure to enter the correct
  conference title from your rights confirmation email}{June 03--05,
  2018}{Woodstock, NY}
\acmISBN{978-1-4503-XXXX-X/2018/06}

\begin{document}

\title{Annie, Are You Okay? How Style- and Context-Based Personalization Shape AI-Assisted Decision-Making}

\author{Hasibur Rahman}
\affiliation{%
  \institution{Northeastern University}
  \city{Boston}
  \state{Massachusetts}
  \country{USA}}
  \email{rahman.has@northeastern.edu}

\author{Benjamin R. Cowan}
\affiliation{%
  \institution{University College Dublin}
  \city{Dublin}
  \country{Ireland}}
  \email{benjamin.cowan@ucd.ie}

\author{Smit Desai}
\authornote{Corresponding author}
\affiliation{
  \institution{Northeastern University}
  \city{Boston}
  \state{Massachusetts}
  \country{USA}}
\email{sm.desai@northeastern.edu}

\renewcommand{\shorttitle}{Effect of Style- and Context-Based Personalization on AI-Assisted Decision Making}
\renewcommand{\shortauthors}{Rahman et al.}

\begin{abstract}
As people turn to generative AI for financial advice, these systems can personalize how they communicate and what they say. Whether these forms of personalization shape decisions differently remains unclear. We conducted a preregistered 2 × 2 between-subjects factorial experiment ($N=240$): participants ranked three comparably viable stocks, discussed them with an AI, and reranked them. Participants perceived both forms of personalization, but only context-based personalization reliably changed ranking behavior: it increased reconsideration and moved rankings toward the AI’s assigned recommendation. Participants felt more influenced without judging the AI as more correct, trustworthy, intelligent, likeable, or high-quality. Those initially farther from its recommendation moved more toward it while judging its advice less correct; exploratory analyses suggest greater susceptibility among lower-expertise participants. These findings show how personalized AI can steer decisions among defensible options with only a minimal evaluative trace, raising concerns for the design and governance of personalized decision support.
\end{abstract}



\begin{teaserfigure}
 \includegraphics[width=\textwidth]{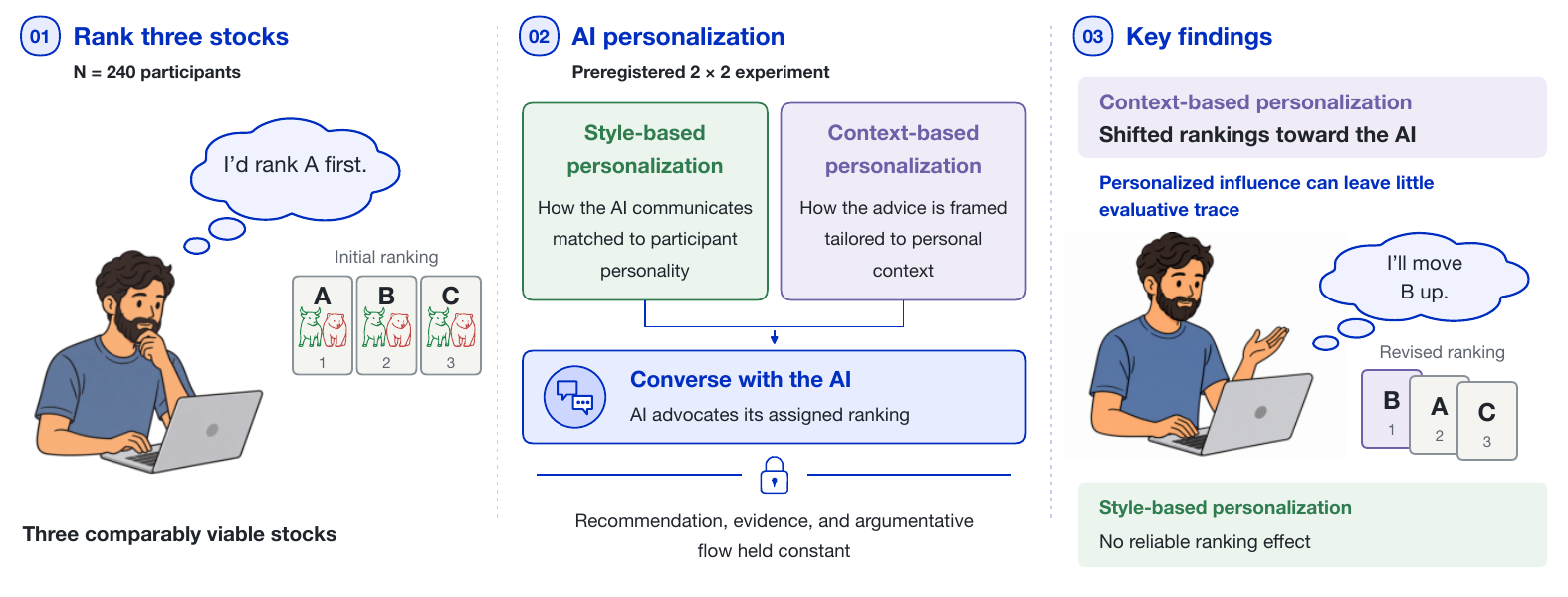}
  \caption{Overview of our study of personalized AI advice ($N=240$). (1) Participants rank three comparably viable stocks. (2) AI conversations independently vary style- and context-based personalization. (3) Reranking and key findings: context-based personalization shifts decisions toward the AI without detectable changes in evaluations of the AI.}
  \Description{Three panels show the study sequence. First, a participant ranks stocks A, B, and C. Second, style-based and context-based personalization are independently varied while the AI advocates an assigned ranking, with recommendation, evidence, and argumentative flow held constant. Third, an illustrative revised ranking places B above A and C. The findings highlight increased movement toward the AI with context-based personalization, no detectable change in AI evaluations, and no reliable ranking effect of style-based personalization.}
  \label{fig:teaser}
\end{teaserfigure}

\maketitle

\section{Introduction}

More than 200 million people ask generative AI for financial advice \emph{every month}. Among Americans who followed such advice, one in five acted immediately, without asking a single follow-up question. Nearly a third said that following such advice hurt their finances.\footnote{\url{https://www.nerdwallet.com/finance/studies/using-ai-for-personal-finances}}


Against this backdrop, in June 2026, ChatGPT introduced a personal finance experience that allows users to connect their financial accounts and save ``financial memories'' to personalize responses.\footnote{\url{https://openai.com/index/personal-finance-chatgpt/}} OpenAI, in its launch announcement, illustrates the feature with two responses to the same request about saving money: one generic, and the other personalized using the user’s location and salary. Both recommend reducing spending and automating savings, but the personalized response uses personal details and preferences to tailor the recommendation's presentation. This shift toward personalized financial advice raises pressing questions—If the recommendation remains the same, can personalization change users' decisions? And does it also change how they perceive the AI giving that advice?

AI can personalize advice in multiple ways \cite{motger2022dialogue,sutcliffe2024survey}. One form is \textit{\facstyle{style-based} personalization}, in which the AI adapts its manner of communication to a user's preferences or personality profile. For example, ChatGPT ask users questions such as, ``Do you like this personality?'', and use their feedback to adapt how it communicates. A second form is \textit{\faccontext{context-based} personalization}, in which the AI uses information about the user to adapt the framing and emphasis of its advice. The ``financial memories'' described above is an example of \faccontext{context-based} personalization.

Prior work suggests that these two mechanisms may influence users differently. Here, persuasive influence refers to a shift in a user's attitude or decision toward the position advanced by the AI. \citet{ruijten2021similarity} found that a dominant AI system was more persuasive but less liked, whereas matching the AI's communication style to the user's personality improved evaluations without improving persuasion. \citet{genc2026bots} likewise found that conversational AI personality\footnote{Following \citet{rahman2026vibecheck}, we use \emph{conversational AI personality} to mean the personality an AI expresses through its linguistic cues and communication style.} affected perceived trustworthiness, competence, and emotional responses without reliably changing user decisions.

Research on profile-personalized messages instead adapts the framing and emphasis of AI messages \cite{matz2017psychological,simchon2024persuasive}. Personalized AI messages can change behavior and sometimes outperform non-personalized alternatives \cite{noar2007tailoring,matz2017psychological,matz2024personalized,salvi2025conversational}. Yet the added benefit of personalization is inconsistent. Political messages generated with AI from demographic and political profiles have performed no better than nontargeted messages \cite{hackenburg2024evaluating,hackenburg2024reply}, while conversational studies identify prompting, post-training, and relevant evidence as stronger drivers of influence \cite{hackenburg2025levers,lin2025persuading}. More broadly, trust can remain stable while reliance rises \cite{salimzadeh2024uncertainty}, and disclosing AI authorship or personalization does not necessarily eliminate influence \cite{matz2024personalized,gallegos2026labeling}.

However, these findings leave the effects of both mechanisms unresolved. It remains unclear how either \facstyle{style-based} or \faccontext{context-based} personalization affects users' decisions and perceptions of the AI, or whether greater movement toward the AI’s recommendation is accompanied by more favorable perceptions of the AI. Existing research cannot answer these questions because they do not isolate the mechanisms. Some pair personalization with other factors, such as opinion alignment \cite{eder2026opinion} or conversational warmth \cite{yazan2026persuade}. Others operationalize personalization by giving the AI access to a participant profile \cite{shen2026puppet}, without separately manipulating how it adapts its communication style and argumentative framing. We therefore lack a controlled test of their separate and combined effects that holds the AI's recommendation, evidence, and argumentative flow constant. Prior research further suggests that this design space is shaped by user-level factors, as users' initial judgments and expertise may affect how readily they can evaluate or contest AI advice \cite{bonaccio2006advice,eichler2025roboadvisor}.

To address these gaps, we conducted a controlled study of personalized AI advice in a preference-sensitive task \cite{elwyn2009dual}. Participants made an initial choice, consulted an AI, and could then keep or revise their choice. In a preregistered $2\times2$ between-subjects experiment ($N=240$), participants ranked three anonymized pharmaceutical stocks, discussed them with an AI advocating a pre-assigned ranking, and reranked them. Real market data supported multiple defensible rankings, with no single objectively correct ordering given the available information. Figure \ref{fig:teaser} summarizes the study design.


Through this study, we examine three research questions: \textbf{RQ1:} \textit{How do \facstyle{style-based} and \faccontext{context-based} personalization affect users’ decisions?} \textbf{RQ2:} \textit{How do \facstyle{style-based} and \faccontext{context-based} personalization affect users’ perceptions of the AI?} and \textbf{RQ3:} \textit{How does users’ initial distance from the AI’s recommendation shape their perceptions and decision changes?} We additionally conduct an exploratory analysis examining how self-reported investment expertise relates to users’ perceptions and susceptibility to AI advice, and whether it moderates the effects of personalization. Together, these questions move from the effects of distinct personalization strategies to the role of initial disagreement to individual differences in susceptibility, offering a comprehensive account of personalized influence in AI-assisted decision-making.

Our findings show that \facstyle{style-} and \faccontext{context-based} personalization have distinct consequences for AI-assisted decision-making. Specifically, with \faccontext{context-based} personalization present, reconsideration increased from 35.0\% to 59.2\%, and the odds of greater movement toward the AI's recommendation approximately doubled ($OR = 2.04$, 95\% CI $[1.23, 3.38]$). Neither form of personalization produced statistically detectable changes in the seven evaluative measures: response quality, perceived correctness, intelligence, likeability, performance trust, moral trust, or willingness to return (all Holm-corrected $p \ge .47$). We contribute:

\begin{itemize}

\item \textbf{A Framework for Studying AI Personalization.} We independently vary how an AI adapts its communication style and tailors its arguments to a user's circumstances. Holding the recommendation, evidence, and conversational sequence constant allows us to test the separate and combined effects of these two forms of personalization.

\item \textbf{Decision Change Without Higher AI Ratings.} \faccontext{Context-based} personalization moved participants toward the AI's recommendation without detectable changes in how they evaluated the AI. Participants recognized the personalization and reported greater influence, raising questions about whether awareness alone is enough to protect users' autonomy.

\item \textbf{Design Implications for Personalized AI Advice.} We propose ways to make the AI's goals and use of personal information visible, give users control over personalization, and assess decision changes alongside users' ratings. We also distinguish safeguards for personalized arguments from the longer-term evaluation needed for communication style.

\end{itemize}

\section{Related Work}

\subsection{Conversational AI in Decision-Making}
\label{sec:rw-aidm}

Most research on AI-assisted decision-making examines a single-turn interaction in which an AI provides a prediction, sometimes with an explanation, and the user decides whether to accept or override it \cite{lai2023science}. CAs instead support multi-turn advice in which the AI can elaborate, answer objections, compare alternatives, and restate its position. These exchanges can improve appropriate reliance over static explanations \cite{ma2025deliberation}, but can also produce overreliance when the AI is wrong \cite{li2025texttotrust,spatharioti2025search}.

\citet{lin2024decisionoriented} described this form of support as \emph{decision-oriented dialogue}. CAs now support decisions from low-stakes reflection to high-stakes choices in life planning \cite{wang2025lifeplanning}, medicine \cite{ramjee2025cataractbot,rajashekar2024clinicalllm}, law \cite{schneiders2025objection}, and public administration \cite{aljuneidi2026stakes}, and the influence of AI advice varies across these domains \cite{ikeda2024chatgptadvice}. We first distinguish users' perceptions of AI from changes in their decisions, then consider personalization and decision settings with multiple defensible answers.

\paragraph{Perceptions of the AI and Changes in Participants' Decisions}
\label{sec:rw-perception}

People's perceptions of AI advisors do not always align with their reliance on its advice. Automation bias names reliance on automated recommendations without adequate scrutiny \cite{parasuraman1997automation}; adding an explanation can increase reliance on correct and incorrect answers alike \cite{kim2025fostering}; trust can remain unchanged while reliance rises under complexity and uncertainty \cite{salimzadeh2024uncertainty}; and explanations can improve understanding without changing reliance \cite{he2023statedaccuracy}. Conversely, radiologists evaluated identical advice less favorably when it carried an AI label, although the label changed neither accuracy nor reliance \cite{gaube2021doasaisay}. Reviews, therefore, caution against using self-reported trust as a proxy for changes in decisions \cite{vereschak2021evaluate,schrills2026questioning}. The same separation appears in persuasive settings. Changes in attitudes and downstream actions may be uncorrelated \cite{hackenburg2026actions}, and users can remain skeptical of individual claims even as the broader argument still moves them \cite{yeo2026deception}. We therefore measure how participants regarded the AI, whether they recognized its influence, and pre-to-post changes in their decisions as separate outcomes. Measuring them separately makes it possible to identify changes in decisions that are not accompanied by corresponding changes in perceptions of the AI.

Susceptibility to AI advice also varies with users' prior judgments and expertise. People often discount advice that falls farther from their initial judgment \cite{harvey1997advice,yaniv2004advice,bonaccio2006advice}, although clinicians have moved toward algorithmic estimates even under strong initial disagreement \cite{palfi2022advicedistance}, and advice consistent with prior beliefs is trusted and accepted more readily \cite{bashkirova2024confirmationbias}. Participants who overestimated their own performance under-relied on an accurate AI \cite{he2023knowing}, and financial literacy predicts responses to robo-advice \cite{eichler2025roboadvisor}. We therefore examine the initial distance between participants' rankings and the AI's assigned ranking (RQ3) and analyze self-reported investment expertise exploratorily.

\subsection{Personalization in Conversational AI}
\label{sec:rw-personalization}

CAs can personalize many aspects of an interaction \cite{motger2022dialogue}, and prior work separates an agent's \emph{persona} from a user's \emph{profile}, although the terms overlap \cite{sutcliffe2024survey}. We instead distinguish how the AI communicates from how it uses participant information. \facstyle{style-based} personalization adapts the AI's communication style to the participant's personality profile, and \faccontext{context-based} personalization adapts the framing and emphasis of its arguments to the participant's demographic and value profile.


\subsubsection{Style-Based Personalization and User--Agent Similarity}
\label{sec:rw-style}

\facstyle{Style-based} personalization draws on similarity-attraction, the tendency to respond more positively to similar others \cite{byrne1971attraction}, which researchers have applied as a design principle in persuasive technology \cite{ruijten2021similarity}. Personality is recognizable in conversational cues, and matching it can affect evaluations. Users rated personality-matched synthesized voices and their content more favorably \cite{nasslee2001synthesized}, readers rated advertisements written for their own Big Five trait more favorably \cite{hirsh2012personalized}, and advertising matched to inferred personality increased clicks and purchases \cite{matz2017psychological}.

Reviews on personalized matching across persuasion research and human-agent interaction find the effects heterogeneous and dependent on which attribute is matched and on the process through which it operates \cite{teeny2021matching, klein2025socialcues}. \citet{ruijten2021similarity} found that a dominant interaction style was more persuasive but less liked, whereas matching the system's style to the user's personality improved perceptions without improving persuasion. In a charitable-giving task, AI personality affected perceived trustworthiness and competence but not donations \cite{genc2026bots}. These findings reinforce the need to distinguish evaluations of an AI from its behavioral influence.

Generative AI allows more flexible style adaptation than earlier systems built on a limited set of designed cues \cite{feine2019taxonomy}. Prompting can control expression of Big Five traits in CAs through reply length, signposting, punctuation, tone, hedging, and related choices \cite{rahman2026vibecheck,serapiogarcia2025psychometric,jiang2023inducing,ramirez2023controlling,jiang2024personallm}. This control makes it possible to revisit personality matching and to examine personality expression and user--agent fit within the same interaction. \citet{rahman2026vibecheck} used this control in a goal-oriented task, found that both the agent's personality expression and its alignment with the user's personality shaped perceptions of the AI, and introduced Trait Modulation Keys (TMK), a prompting framework that jointly controls Big Five profiles and their communication styles; we use TMK for \facstyle{style-based} personalization (Section~\ref{sec:style-layer}). 

Prior work, however, has more often varied an AI's personality than adapted its communication to an individual participant, and has focused more on evaluating user perceptions than on its impact on decisions. To address this, we test whether \facstyle{style-based} personalization affects participants' decisions (RQ1), their perceptions of the AI (RQ2), or both.

\subsubsection{Context-Based Personalization and Tailored Persuasion}
\label{sec:rw-context}

\faccontext{Context-based} personalization adapts what an AI says based on information about the user. Such personalization raises persuasive relevance: arguments framed around recipients' moral values shift attitudes more than generic framings \cite{feinberg2015gulf,feinberg2019moral}, and across 57 studies of print health materials personalized to readers' demographics, behavior, and psychological attributes, personalization gave an advantage in changing health behaviors \cite{noar2007tailoring}. AI can generate such messages from user profiles. Profile-personalized messages outperformed non-personalized ones even when the personalization was disclosed \cite{matz2024personalized}, and personality-matched political advertisements were rated more persuasive than mismatched ones \cite{simchon2024persuasive}.

Evidence that such personalization adds to AI persuasion remains mixed. Political messages generated from demographic and political profiles were no more persuasive than untargeted ones \cite{hackenburg2024evaluating,hackenburg2024reply}. Conversational findings are mixed as well. In short online debates, AI personalized to participants' sociodemographic profiles shifted post-debate agreement more often than human debaters \cite{salvi2025conversational}, while larger-scale studies found stronger effects from post-training and prompting than from personalization, and identified relevant facts and evidence as central to persuasion \cite{hackenburg2025levers,lin2025persuading}. Context-personalized dialogues with AI reduced conspiracy beliefs by about 20\%, with the effect lasting 2 months \cite{costello2024durably}.

The decision domain and contents of the user profile may explain part of this variation. Most prior research on personalized persuasion concerns political attitudes, but persuasive effects vary across communication contexts \cite{holbling2025metaanalysis} and decision domains \cite{ikeda2024chatgptadvice}.  Personal attributes may bear directly on the decision rather than serving only as targeting variables, and values connect an argument to a recipient's priorities \cite{feinberg2015gulf,feinberg2019moral}. Our \faccontext{context-based} personalization therefore adapts the framing and emphasis of the argument to the participant's demographic and value profile (Section \ref{sec:context-layer}), leaving the assigned recommendation, evidence, and argumentative flow unchanged.

\subsection{Decision-Making with Multiple Defensible Answers}
\label{sec:rw-finance}

Following \citet{elwyn2009dual}, we describe decisions as \emph{preference-sensitive} when several options are defensible, and the preferred choice depends on how individuals weigh their benefits and risks. Evaluating AI advice in these settings requires distinguishing movement toward its recommendation from improvement in decision quality~\cite{guo2024reliance}. Such tasks allow us to examine whether an AI recommendation changes a reasoned judgment without assuming that agreement with the AI is objectively correct. The absence of a single correct answer defines the decision setting; controlled comparisons are still needed to identify the effects of personalization.

Investment selection provides one such setting. Portfolio choice balances expected return against the risk an investor will accept \cite{markowitz1952portfolio}, and investors can weigh growth, valuation, and risk differently~\cite{elliott2023asymmetric}. HCI studies have used robo-advised life-insurance choices \cite{bertrand2023roboadvised}, portfolios built from real ETFs~\cite{reicherts2025think}, and real-stock rankings after conversations with an AI advisor \cite{takayanagi2025advisors}. Related work examines explanations in simulated financial advice \cite{bendavid2021xai}, trust in personal-investing robo-advisors \cite{cooney2026roboadvisors}, and the acceptability of delegating investments to algorithms \cite{niszczota2020roboinvestment}. This literature establishes the investment selection task as a setting for studying responses to AI advice.

\subsection{Isolating Style- and Context-Based Personalization}

Across these literatures, \facstyle{style-based} and \faccontext{context-based} personalization are rarely varied independently.  Research on user--agent similarity typically varies how an agent communicates, such as an extraverted or introverted voice \cite{nasslee2001synthesized}, a dominant or submissive interaction style \cite{ruijten2021similarity}, or a fast and proactive versus polite and accommodating conversational style \cite{kostric2025tailor}, while the content stays the same; psychological-targeting research instead changes a message's framing and emphasis \cite{matz2017psychological,matz2024personalized,simchon2024persuasive,hackenburg2024evaluating}. Recent work has begun to examine related dimensions together. Researchers have crossed an assistant's personality with opinion alignment \cite{eder2026opinion} and contextualization with warmth \cite{yazan2026persuade}, and a study of everyday advice-seeking conversations bundled demographics, values, and personality into one personalization factor \cite{shen2026puppet}. These designs examine multiple dimensions of an interaction, but none independently manipulates the manner of communication and the framing of the argument.

Motivated by these gaps, we independently vary both forms in a preference-sensitive investment-ranking task. We test their effects on decisions (RQ1) and perceptions of the AI (RQ2), and examine associations with initial distance from the AI's assigned ranking (RQ3).

\section{Study Method}

To test whether personalizing how an AI communicates and how it frames its arguments changes users’ perceptions and investment decisions, we conducted an Institutional Review Board-approved and preregistered experiment ($N=240$). The study was built around three design requirements. First, the decision task needed to support multiple defensible answers so that influence could be studied without presuming that the AI held a uniquely correct position. Second, \facstyle{style-} and \faccontext{context-based} personalization needed to vary independently while other features of the AI interaction remained controlled. Third, the design needed to distinguish changes in participants’ decisions from their perceptions of the AI.

We used investment ranking as an instance of this decision setting (Section \ref{sec:rw-finance}), selecting three anonymized pharmaceutical stocks from real market data as comparably viable long-term investments. Each company was presented with a bull case—an argument for investing—and a bear case—an argument against investing—so that participants had evidence supporting different conclusions. As reported in Section \ref{sec:random}, participants’ initial choices included all six possible rankings, with no stock or ordering emerging as an obvious favorite. The AI advocated one of these six rankings. Its ranking was assigned in advance, counterbalanced within each experimental cell, and held independent of the participant’s initial choice. Initial agreement or disagreement with the AI could therefore vary naturally rather than being imposed as part of the personalization manipulation.

To isolate the two forms of personalization, we used a $2 \times 2$ between-subjects factorial design with 60 participants per cell. \facstyle{Style-based} personalization was either absent or present, and \faccontext{context-based} personalization was independently either absent or present (see Table \ref{tab:factorial-design}), producing four design cells: \baseline{Baseline} (Style: absent, Context: absent), \style{Style-only} (Style: present, Context: absent), \context{Context-only} (Style: absent, Context: present), and \combined{Combined} (Style: present, Context: present). \facstyle{Style-based} personalization adapted the AI’s manner of communication to the participant’s personality profile, whereas \faccontext{context-based} personalization adapted the framing and emphasis of its arguments to the participant’s demographic and value profile. The AI’s assigned recommendation, underlying evidence, argumentative flow, conversational interface, and study procedure were otherwise held constant.

Behavioral outcomes were measured before and after the conversation, adding Time as a within-participant measurement factor to the randomized $2 \times 2$ design. We assessed behavioral influence through pre-to-post changes in participants’ rankings, investment likelihoods, and confidence. Perceptions of the AI and manipulation checks were collected only after the conversation. We additionally examined whether participants’ responses varied with the initial distance between their ranking and the AI’s assigned ranking.

Section~\ref{sec:task} describes the stock selection task, Section~\ref{sec:conditions} defines the experimental cells, and Section~\ref{sec:prompt-design} explains their prompt implementation. We then describe the measures, procedure, participants, and analysis strategy.

\begin{center}
\begin{minipage}{\linewidth}
\captionof{table}{The $2 \times 2$ between-subjects factorial design ($N=240$). Style-based and context-based personalization were independently absent or present, with 60 participants per experiment cell.}
\label{tab:factorial-design}
\centering
\small
\renewcommand{\arraystretch}{1.2}
\begin{tabular}{@{}lcc@{}}
\toprule
 & \multicolumn{2}{c}{\textbf{Style-based personalization}} \\
\cmidrule(l){2-3}
\textbf{Context-based personalization} & Absent & Present \\
\midrule
Absent & Baseline ($n=60$) & Style-only ($n=60$) \\
Present & Context-only ($n=60$) & Combined ($n=60$) \\
\bottomrule
\end{tabular}
\end{minipage}
\end{center}

\subsection{Designing the Stock Selection Task}
\label{sec:task}
To assess the effect of personalization, we asked participants to rank three companies, labeled Company A, Company B, and Company C, by their attractiveness as long-term investments. Our task builds on prior HCI studies of investment decision support, which have asked participants to construct portfolios from real exchange-traded funds \cite{reicherts2025think} or rank real stocks after a conversation with an AI advisor \cite{takayanagi2025advisors}. Unlike tasks scored against an expert ranking \cite{takayanagi2025advisors}, our task specified no correct ordering: all three were comparably viable investments whose ranking could depend on participants' priorities. Ranking rather than choosing one stock lets us measure partial movement toward or away from the AI's ranking (Section \ref{sec:behavioral-measures}).


We selected three pharmaceutical companies in three stages. First, using the Morningstar industry report\footnote{\url{https://www.morningstar.com/}} and Robinhood,\footnote{\url{https://robinhood.com/}} we retained the 30 largest publicly traded U.S. pharmaceutical companies by market capitalization. Restricting candidates to traditional drug manufacturers ensured comparable fundamentals. All subsequent screening used Robinhood data. Second, we retained the five stocks with the largest positive net buying values over the preceding month, reflecting share purchases net of sales. Third, using Robinhood's reported market capitalizations, we selected three companies spanning a range of sizes. We used positive net buying as a shared indicator of investor demand and variation in company size to represent different growth profiles, supporting defensible rankings for different long-term preferences.

We obtained the selected companies' market data from the NYSE\footnote{\url{https://www.nyse.com/}} on June 2, 2026. The data included opening price, volume, market capitalization, daily and annual lows and highs, trailing 12-month earnings per share, price-to-earnings ratio, and price histories over six windows from 1 month to the full record. We obtained each bull and bear case from SoFi\footnote{\url{https://www.sofi.com/invest/}} on the same date. We labeled the companies A, B, and C and removed specific drug names from the bull and bear cases.

We used real company data to preserve ecological validity \cite{raggetti2017trading, vanbrussel2024investors}, while presenting the companies as fictional to limit the influence of participants' prior views. Experienced investors thus encountered plausible fundamentals and arguments. Section \ref{sec:random} confirms that participants treated all three as viable options.

\begin{figure*}[p]
\centering
\begin{subfigure}[t]{\textwidth}
  \centering
  \includegraphics[width=\textwidth]{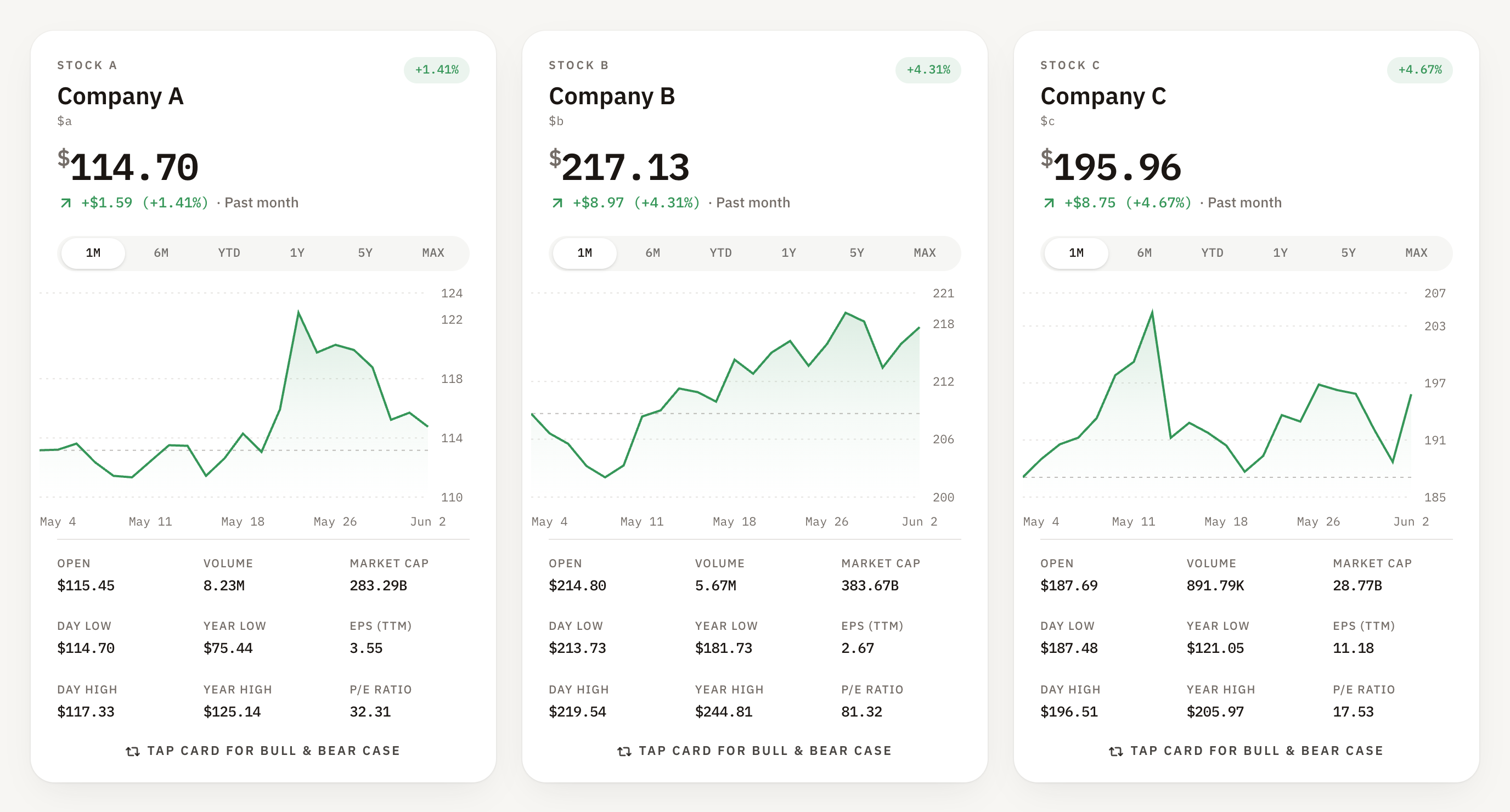}
  \caption{\textbf{Front of the cards.} Each card gave the current price, a line chart of the price history over a selectable window, and a grid of nine fundamentals.}
  \label{fig:app-cards}
\end{subfigure}

\vspace{8pt}

\begin{subfigure}[t]{\textwidth}
  \centering
  \includegraphics[width=\textwidth]{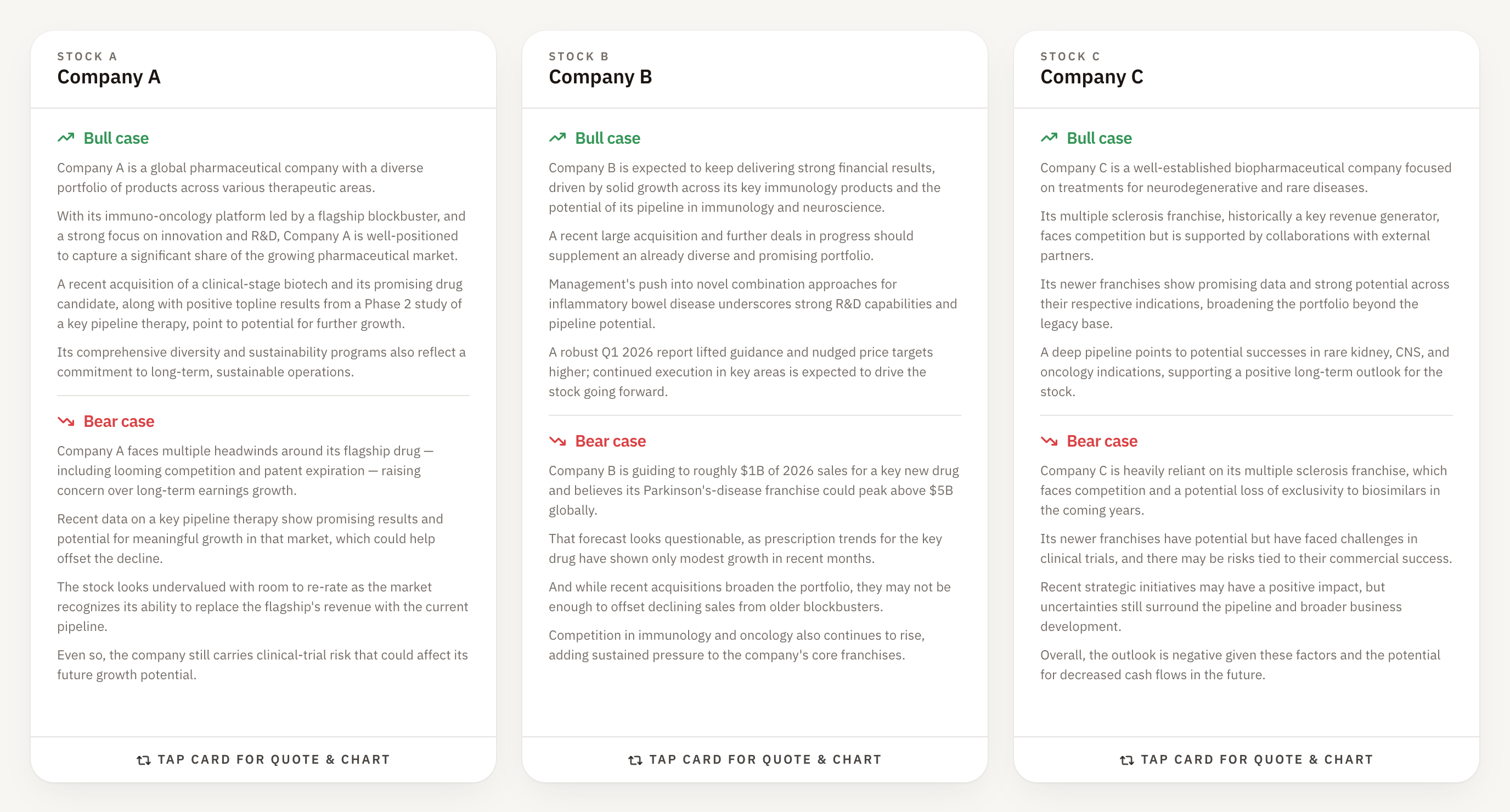}
  \caption{\textbf{Back of the cards.} Tapping a card turned it over to a written bull case and bear case, so every company carried an argument on both sides.}
  \label{fig:app-cases}
\end{subfigure}
\caption{The stock selection task interface. Each company appeared on its own card, and that card carried all the information participants had about the company: the market data on the front and the arguments on both sides on the back.}
\label{fig:task-ui}
\Description{Two rows of screenshots. (a) Three cards side by side for Company A, Company B, and Company C. Each card shows a ticker, a price, the change over the past month, a row of tabs for the time window, a line chart of the price history, and a grid of fundamentals covering opening price, volume, market capitalization, day low and high, year low and high, earnings per share, and price-to-earnings ratio. A line at the bottom of each card reads ``Tap card for bull and bear case.'' (b) The same three cards turned over, each showing a paragraph-length bull case in green followed by a bear case in red, with a line at the bottom reading ``Tap card for quote and chart.''}
\end{figure*}

All companies used the same card format (Figure \ref{fig:task-ui}). The front showed its current price, change, and line chart over six selectable windows, and the same fundamentals (Figure \ref{fig:app-cards}); the back showed its bull and bear cases (Figure \ref{fig:app-cases}). Bull and bear cases supported any ordering; random card order avoided implying a ranking.

\subsection{Experimental Design}
\label{sec:conditions}
We randomly assigned participants to four experiment cells in a $2 \times 2$ between-subjects factorial design ($n=60$ per experimental cell; Table~\ref{tab:factorial-design}). \facstyle{Style-based} personalization adapted the AI's communication to the participant's Big Five personality profile through reply length, signposting, punctuation, tone, and hedging. \faccontext{Context-based} personalization adapted the framing and emphasis of stock arguments to the participant's demographic and value profile. Crossing the two factors yielded these experiment cells:

\begin{itemize}
    \item \baseline{Baseline (Style absent, Context absent).} The AI used plain declaratives and a flat tone, without a demographic or value profile to tailor its arguments.
    \item \style{Style-only (Style present, Context absent).} The AI adapted its communication to the participant's personality, without a demographic or value profile to tailor its arguments.
    \item \context{Context-only (Style absent, Context present).} The AI used plain declaratives and a flat tone while connecting stock arguments to the participant's circumstances and values.
    \item \combined{Combined (Style present, Context present).} The AI adapted its communication to the participant's personality and its argument framing to their circumstances and values.
\end{itemize}

All four experiment cells used the same company evidence, ranking-assignment procedure, argumentative sequence, conversational interface, and study procedure. In every experiment cell, the AI received participants' initial rankings, investment likelihoods, and rationales. Thus, the baseline also responded to participants' task-specific input; the two factors varied personality-based communication and demographic- and value-based framing. Section~\ref{sec:prompt-design} describes their prompt implementation.

\paragraph{Assigning the AI's Reference Ranking}
\label{sec:simulated-ranking}
To measure movement toward the AI, we assigned it a ranking to advocate, following studies that assign AI systems a position and assess users' subsequent attitudes or choices~\cite{jakesch2023opinionated,williamsceci2026biased,salvi2026commercial,werner2024steering,chiang2024devilsadvocate,shen2026puppet}. Within each experiment cell, we rotated through all $3!=6$ rankings in participant-arrival order, assigning each ten times. The application fixed the ranking when participants reached the conversation screen, independently of their initial rankings or subsequent messages. The AI presented this ranking as its own judgment throughout the conversation. This counterbalancing separated the advocated position from personalization, while allowing initial agreement with the AI to vary naturally. Section~\ref{sec:random} reports balance checks; debriefing disclosed the advance assignment (Section~\ref{sec:procedure}).

\subsection{Prompt Design}
\label{sec:prompt-design}
\begin{figure}[t]
\centering
\includegraphics[scale=.8]{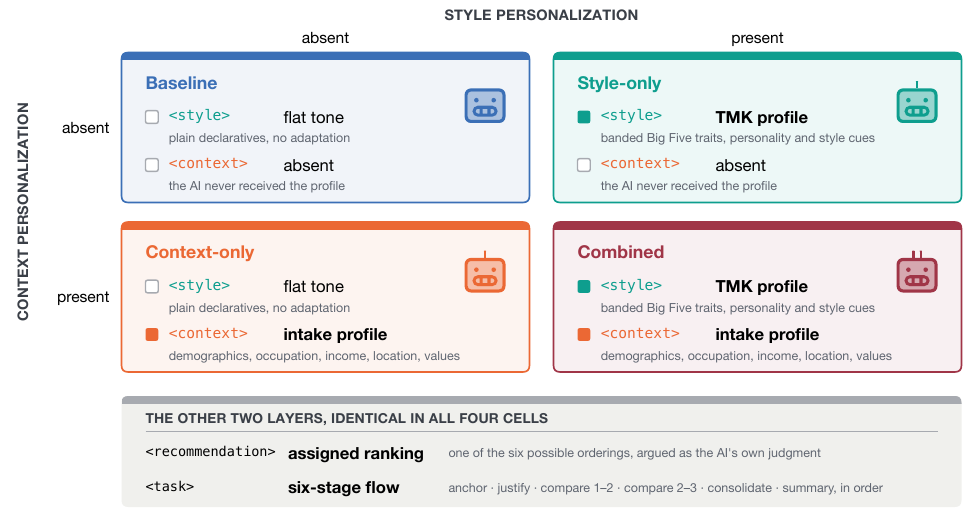}
\caption{\textbf{Independent personalization factors implemented through prompt layers.} Each experimental cell sets \lyrstyle{} and \lyrcontext{}; recommendation instructions and task rules are shared. The assigned ranking is counterbalanced within each experimental cell.}
\label{fig:conditions}
\Description{A two-by-two grid of four cards, one per experimental cell, on axes for style-based personalization (absent, present) across the top and context-based personalization (absent, present) down the side. Baseline sits top left with the style layer set to flat tone and the context layer absent; Style-only, top right, sets the style layer to the TMK profile; Context-only, bottom left, sets the context layer to the intake profile; Combined, bottom right, sets both. Each card also shows a robot whose antennas count the personalized layers: none for Baseline, one each for Style-only and Context-only, and two for Combined. A gray band below the grid, headed the other two layers identical in all four experimental cells, gives the recommendation layer as an assigned ranking, one of the six possible orderings argued as the AI's own judgment, and the task layer as a six-stage flow of anchor, justify, compare one and two, compare two and three, consolidate, and summary, in order.}
\end{figure}

We used GPT-5.4 in all four experimental cells.\footnote{Accessed through OpenRouter (\url{https://openrouter.ai/}), with reasoning effort set to medium.} A \emph{prompt layer} is a group of instructions within the same system prompt that governs one aspect of the AI's responses. In our study, the system prompt contained four layers: recommendation, task, style, and context (Figure~\ref{fig:conditions}). The \lyrrec{} layer specified the assigned ranking and instructed the AI to present it as its own judgment. The \lyrtask{} layer specified the argumentative sequence, turn-level rules, and evidence constraints. The \lyrstyle{} and \lyrcontext{} layers implemented the respective personalization factors using separate personality and demographic/value profiles. Recommendation instructions and task rules were identical across experimental cells; the assigned ranking varied according to the counterbalancing procedure. Every prompt included the company profiles, bull and bear cases, and participants' initial rankings, investment likelihoods, and rationales. The full prompts are available in the supplementary materials.

\subsubsection{Shared Argumentative Sequence}
\label{sec:convo-flow}
Because persuasive effects depend on implementation and communication context~\cite{holbling2025metaanalysis}, \lyrtask{} prescribed the same six stages across experimental cells:

\begin{enumerate}
    \item \textbf{Anchor.} Open with the full ranking and one evidence-based clause per stock to establish a reference point~\cite{tversky1974uncertainty}. Foreground the bull case for a stock the AI ranked higher than the participant did and the key risk of their favored stock, or affirm an identical ranking. Acknowledge their rationale~\cite{cialdini2007influence} and ask why they favored their top pick.
    \item \textbf{Justify.} Support the top-ranked stock with issue-relevant evidence and numbers, following the central route of the Elaboration Likelihood Model~\cite{petty1986elm}.
    \item \textbf{Compare first versus second.} Concede a strength of the second-ranked stock, then defend the first, using two-sided argumentation~\cite{allen1991twosided,okeefe1999twosided}.
    \item \textbf{Compare second versus third.} Repeat the two-sided comparison for the lower pair.
    \item \textbf{Consolidate criteria.} Restate the ranking in terms of current financial support, long-term growth, and risk concentration, using emphasis framing~\cite{entman1993framing}.
    \item \textbf{Summary.} Restate the ranking with a one-line rationale per stock.
\end{enumerate}

The prompt required one stage per reply, in sequence, with brief answers to participants' questions before returning to the flow. It restricted numbers and bull or bear points to the displayed company profiles. It also required periodic reminders that the decision remained the participant's own, an autonomy-preserving formulation associated with lower reactance~\cite{carpenter2013free}, and prohibited asking participants to revise their rankings. Section~\ref{sec:llm-judge} assesses adherence to these constraints.

\subsubsection{Style Instructions}
\label{sec:style-layer}
With style present, \lyrstyle{} used the Trait Modulation Keys (TMK) of \citet{rahman2026vibecheck}, which map low, medium, or high levels of each Big Five trait to communication cues. We converted participants' Mini International Personality Item Pool scores (Section~\ref{sec:intake-measures}) into five trait means and banded each trait. Because the Mini-IPIP has no published norms and the International Personality Item Pool recommends local norms,\footnote{\url{https://ipip.ori.org/}} we fitted one-dimensional $k$-means clusters to a publicly available, demographically matched dataset of 250 participants. Silhouette scores were 0.57--0.64; across 1{,}000 bootstrap resamples, 91--98\% of participants retained their band. The banded profile populated \lyrstyle{}. For high extraversion, the instructions read:

\begin{quote}\small\itshape
You are active, energetic, and talkative. You are extraverted, bold, and assertive in conversations. Insert exclamation marks for enthusiasm, unicode emojis or interjections [...].
\end{quote}

With style absent, \lyrstyle{} prescribed plain declaratives and a flat tone.

\subsubsection{Context Instructions}
\label{sec:context-layer}
With context present, \lyrcontext{} supplied age group, gender, ethnicity, education, occupation, location, household income, household status, and four higher-order Schwartz value dimensions (Section~\ref{sec:intake-measures}). It instructed:

\begin{quote}\small\itshape
Every reply ties the argument to at least one concrete fact of their life, addressed as ``you / your'' --- their work, their age and life stage, living single or with others, their income band, or where they live. Name the fact and connect it to the stock point in the same breath.
\end{quote}

The instructions standardized which attributes entered each stage: the strongest value and occupation set the anchor's leading criterion; occupation and education shaped the justification, analogy, and numerical density; income and household set the stakes in the first comparison; location and age set the time horizon in the second; and closing stages drew on the full profile. Values appeared as plain priorities (e.g., ``you'd rather own something that lasts than chase a quick flip''). These rules standardized the extent and timing of personalization while preserving the assigned ranking, stock facts, and argumentative sequence. With context absent, the prompt supplied no demographic or value profile.

\subsection{Measures}
We describe four sets of measures: intake measures, behavioral measures, perception measures, and manipulation-check indices. Table \ref{tab:measures} lists every measure with its range, when it was collected, and the research question it serves.

\begin{table}[t]
\centering
\footnotesize
\setlength{\tabcolsep}{6pt}
\renewcommand{\arraystretch}{1.1}
\caption{Measures, when they were collected, and where they are used. Pre and Post refer to the behavioral battery before and after the AI conversation; $d_1$ is the initial distance between the participant's ranking $r_1$ and the AI's assigned ranking $a$; Expl.\ is the exploratory expertise analysis. Definitions are given in the sections named in each group heading.}
\label{tab:measures}
\begin{tabular}{@{}llll@{}}
\toprule
Measure & Range & Collected & Used in \\
\midrule
\rowcolor{rowshade} \multicolumn{4}{@{}l}{\textit{Intake measures} (Section \ref{sec:intake-measures})} \\
Demographics and circumstances & Categorical & Intake & \lyrcontext{} input \\
Personal values (SSVS) & $-1$ to 8 & Intake & \lyrcontext{} input \\
Big Five personality (Mini-IPIP) & 1--5 & Intake & \lyrstyle{} input \\
Investment experience & 4 levels & Intake & Expl. \\
\midrule
\rowcolor{rowshade} \multicolumn{4}{@{}l}{\textit{Behavioral battery} (Section \ref{sec:behavioral-measures})} \\
Stock ranking $r_1$, $r_2$ & 6 orderings & Pre, Post & Basis of ranking measures \\
Investment likelihood $L_s$ & 0--100 & Pre, Post & Basis of likelihood measures \\
Confidence $C$ & 0--100 & Pre, Post & Basis of confidence change \\
Rationale & Text & Pre, Post & AI prompt input \\
\midrule
\rowcolor{rowshade} \multicolumn{4}{@{}l}{\textit{Derived decision measures}} \\
Initial distance $d_1 = F(r_1, a)$ & 0, 2, 4 & Pre & RQ3 (moderator) \\
Reconsideration & 0/1 & Pre, Post & RQ1, RQ3, Expl. \\
Movement toward AI & $-(4 - d_1)$ to $d_1$ & Pre, Post & RQ1, RQ3, Expl. \\
Rate of change & $-1$ to 1 & Pre, Post & RQ3 check \\
Ranking change (footrule) & 0, 2, 4 & Pre, Post & RQ1, RQ3, Expl. \\
Ranking change (Kendall) & 0--3 & Pre, Post & RQ1, RQ3, Expl. \\
Likelihood shift toward AI & $-200$ to 200 & Pre, Post & RQ1, RQ3, Expl. \\
Likelihood change, AI-top stock & $-100$ to 100 & Pre, Post & RQ1, RQ3, Expl. \\
Total absolute likelihood change & 0--300 & Pre, Post & RQ1, RQ3, Expl. \\
Confidence change & $-100$ to 100 & Pre, Post & RQ1, RQ3, Expl. \\
\midrule
\rowcolor{rowshade} \multicolumn{4}{@{}l}{\textit{Perception measures} (Section \ref{sec:perception-measures})} \\
Response quality & 1--7 & Post & RQ2, RQ3, Expl. \\
Rightness judgment & 1--7 & Post & RQ2, RQ3, Expl. \\
Return likelihood & 1--7 & Post & RQ2, RQ3, Expl. \\
Likeability (Godspeed) & 1--5 & Post & RQ2, RQ3, Expl. \\
Perceived intelligence (Godspeed) & 1--5 & Post & RQ2, RQ3, Expl. \\
Performance trust (MDMT) & 0--7 & Post & RQ2, RQ3, Expl. \\
Moral trust (MDMT) & 0--7 & Post & RQ2, RQ3, Expl. \\
Perceived influence & 1--7 & Post & RQ2, RQ3, Expl. \\
\midrule
\rowcolor{rowshade} \multicolumn{4}{@{}l}{\textit{Manipulation checks} (Section \ref{sec:manipulation-checks})} \\
Style index & 1--7 & Post & Manipulation validation \\
Context index & 1--7 & Post & Manipulation validation \\
\bottomrule
\end{tabular}
\end{table}

\subsubsection{Intake Measures}
\label{sec:intake-measures}
The intake questionnaire covered demographics, AI use, investment experience, and current stock holdings; the Short Schwartz Value Survey \cite{lindeman2005ssvs}, aggregated into Schwartz's four higher-order dimensions \cite{schwartz1992universals}; and the 20-item Mini International Personality Item Pool (Mini-IPIP) \cite{donnellan2006miniipip} for the Big Five. Demographics and values supplied \lyrcontext{}; Big Five scores supplied \lyrstyle{}.

\subsubsection{Behavioral Measures}
\label{sec:behavioral-measures}
Before and after the conversation, participants ranked stocks from most (1) to least (3) attractive as long-term investments, rated investment likelihood for each and ranking confidence, and explained their ranking, following pre--post advice-taking designs \cite{bonaccio2006advice} and AI-reliance research \cite{raees2026reliance}.

Let $r_1$, $r_2$, and $a$ denote participants' pre-conversation, post-conversation, and AI-advocated rankings, respectively. We use two established rank metrics \cite{diaconis1977footrule} used in HCI to compare predicted and user-adjusted rankings \cite{sertkan2020touristic} and stakeholder and decision-support-system rankings \cite{shrestha2023help}. The Spearman footrule \cite{spearman1906footrule} sums each stock's absolute change in position,
\begin{equation*}
F(x, y) = \sum_{s \in \{A, B, C\}} \bigl| \mathrm{pos}_x(s) - \mathrm{pos}_y(s) \bigr|,
\end{equation*}
and the Kendall tau distance \cite{kendall1938tau} counts stock pairs in opposite relative order,
\begin{equation*}
K(x, y) = \#\bigl\{ \{s, s'\} : \bigl(\mathrm{pos}_x(s) - \mathrm{pos}_x(s')\bigr)\bigl(\mathrm{pos}_y(s) - \mathrm{pos}_y(s')\bigr) < 0 \bigr\}.
\end{equation*}
Over three stocks, $F \in \{0, 2, 4\}$ and $K \in \{0, 1, 2, 3\}$: $F$ counts how far stocks traveled, $K$ how many pairwise preferences flipped.

\emph{Initial distance}, $d_1 = F(r_1, a)$, measures pre-exposure disagreement with the AI and moderates RQ3. \emph{Reconsideration} is 1 if $r_2 \ne r_1$ and 0 otherwise. \emph{Movement toward the AI} is $d_1 - F(r_2, a)$, positive when participants ended closer to the AI, and \emph{ranking change} is $F(r_1, r_2)$ or $K(r_1, r_2)$, movement regardless of direction; these generalize answer switching and agreement with the AI \cite{yin2019accuracy, raees2026reliance} to orderings. With $\Delta$ denoting post minus pre and $t$ and $b$ the stocks the AI ranked first and last, \emph{likelihood shift toward the AI} is $\Delta L_t - \Delta L_b$, \emph{likelihood change for the AI-top stock} is $\Delta L_t$, \emph{total absolute likelihood change} is $\sum_s |\Delta L_s|$, and \emph{confidence change} is $\Delta C$.

Movement toward the AI cannot exceed $d_1$; initially agreeing participants ($d_1 = 0$) cannot move closer. We compute a \emph{rate of change}, $\bigl(d_1 - F(r_2, a)\bigr) / d_1$ for $d_1 > 0$, the fraction of available room used, as a robustness check (Appendix \ref{sec:rq3-rate}). It applies the weight-of-advice normalization, dividing movement toward an advisor by the initial gap for human and algorithmic advice \cite{harvey1997advice,yaniv2004advice,logg2019algorithm}, to complete rankings by replacing scalar differences with footrule distances; it parallels the switch fraction of \citet{yin2019accuracy}, whose denominator includes only initially disagreeing participants.

For example, a participant starting at $r_1 = (B, A, C)$ when the AI advocates $a = (A, B, C)$ has $d_1 = F(r_1, a) = 2$: two stocks are each one position away. Ending at $r_2 = (A, B, C)$ gives a final distance of $F(r_2, a) = 0$, movement toward the AI of $2 - 0 = 2$, and a rate of change of $2 / 2 = 1$, using all available room. Swapping the top two stocks yields ranking changes of $F(r_1, r_2) = 2$ and $K(r_1, r_2) = 1$. If the participant's likelihood for stock $A$ rose by 20 points, fell by 10 for stock $C$, and was unchanged for stock $B$, the likelihood shift toward the AI is $20 - (-10) = 30$, the likelihood change for the AI-top stock is 20, and the total absolute likelihood change is 30.

\subsubsection{Perception Measures}
\label{sec:perception-measures}
We assembled eight perception measures spanning \citet{wei2023bot}'s categories of interaction, ability, and humanness. We report Cronbach's $\alpha$ for every multi-item measure. Three single items adapted from \citet{cheng2026sycophantic} capture \emph{response quality}, \emph{rightness judgment} of the AI's ranking, and \emph{return likelihood} for similar questions. Two Godspeed scales \cite{bartneck2009godspeed} each average five semantic-differential pairs: \emph{likeability} (e.g., unfriendly to friendly) and \emph{perceived intelligence} (e.g., incompetent to competent). Two subscales of the Multi-Dimensional Measure of Trust \cite{ullman2019trust}, averaged over rated items, capture \emph{performance trust} (eight reliable and capable adjectives) and \emph{moral trust} (eight ethical and sincere adjectives). \emph{Perceived influence} averages three agreement statements assessing AI influence on the final decision, effects on ranking, and consideration in deciding.

\subsubsection{Manipulation Questions}
\label{sec:manipulation-checks}
Four agreement statements formed two indices. The \emph{style index} averages two items asking whether the AI's communication felt similar to the participant's own and tailored to them; the \emph{context index} averages two items assessing whether its recommendations appeared tailored to the participant and drew on information about them.

\subsection{Procedure}
\label{sec:procedure}

\begin{figure*}[!ht]
\centering
\includegraphics[scale=.8]{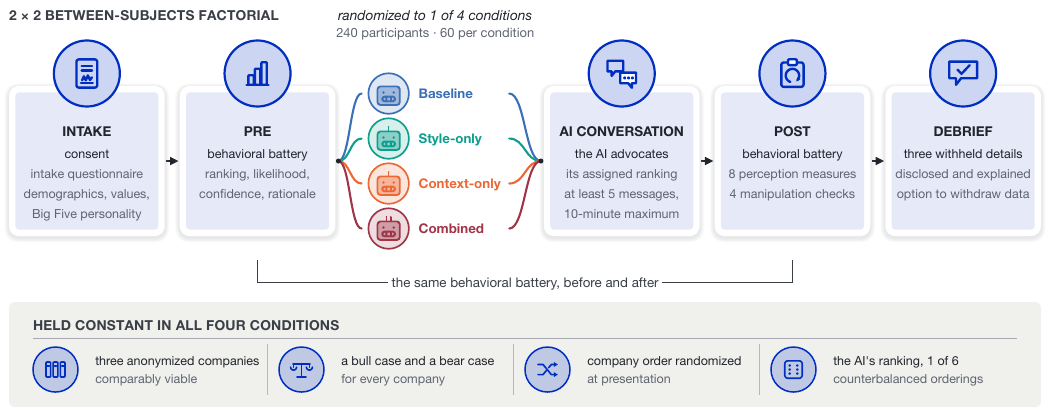}
\caption{\textbf{Study procedure and controls.} Participants completed the intake and the first behavioral battery before random assignment to one of four experimental cells, conversed with an AI advocating a counterbalanced ranking, repeated the battery, and received a debriefing. Company information, evidence structure, presentation-order randomization, and ranking counterbalancing remained constant across experimental cells.}
\Description{A horizontal flow diagram of the study. Participants move through intake, a pre-conversation behavioral assessment, random assignment to the Baseline, Style-only, Context-only, or Combined experimental cell, an AI conversation, a post-conversation behavioral and perception assessment, and debriefing. A bracket links the pre- and post-conversation assessments to indicate that participants completed the same behavioral battery twice. A band below lists controls shared across experimental cells: three anonymized, comparably viable companies; bull and bear cases for each company; randomized company presentation order; and one of six counterbalanced AI rankings.}
\label{fig:procedure}
\end{figure*}

Participants entered the custom study application through Prolific\footnote{\url{https://www.prolific.com/}} and completed four phases: intake, pre-conversation assessment, AI conversation, and post-conversation assessment (Figure \ref{fig:procedure}).

\textbf{Intake.} Participants reviewed the study information, provided consent, and completed the intake questionnaire (Section \ref{sec:intake-measures}), which included demographic, value, and personality profile. When enabled, personalization drew on these responses.

\textbf{Pre-conversation assessment.} Participants read the task instructions, reviewed the three randomly ordered company profiles, and completed the behavioral battery (Section \ref{sec:behavioral-measures}).

\begin{figure*}[t]
\centering
\includegraphics[width=\textwidth]{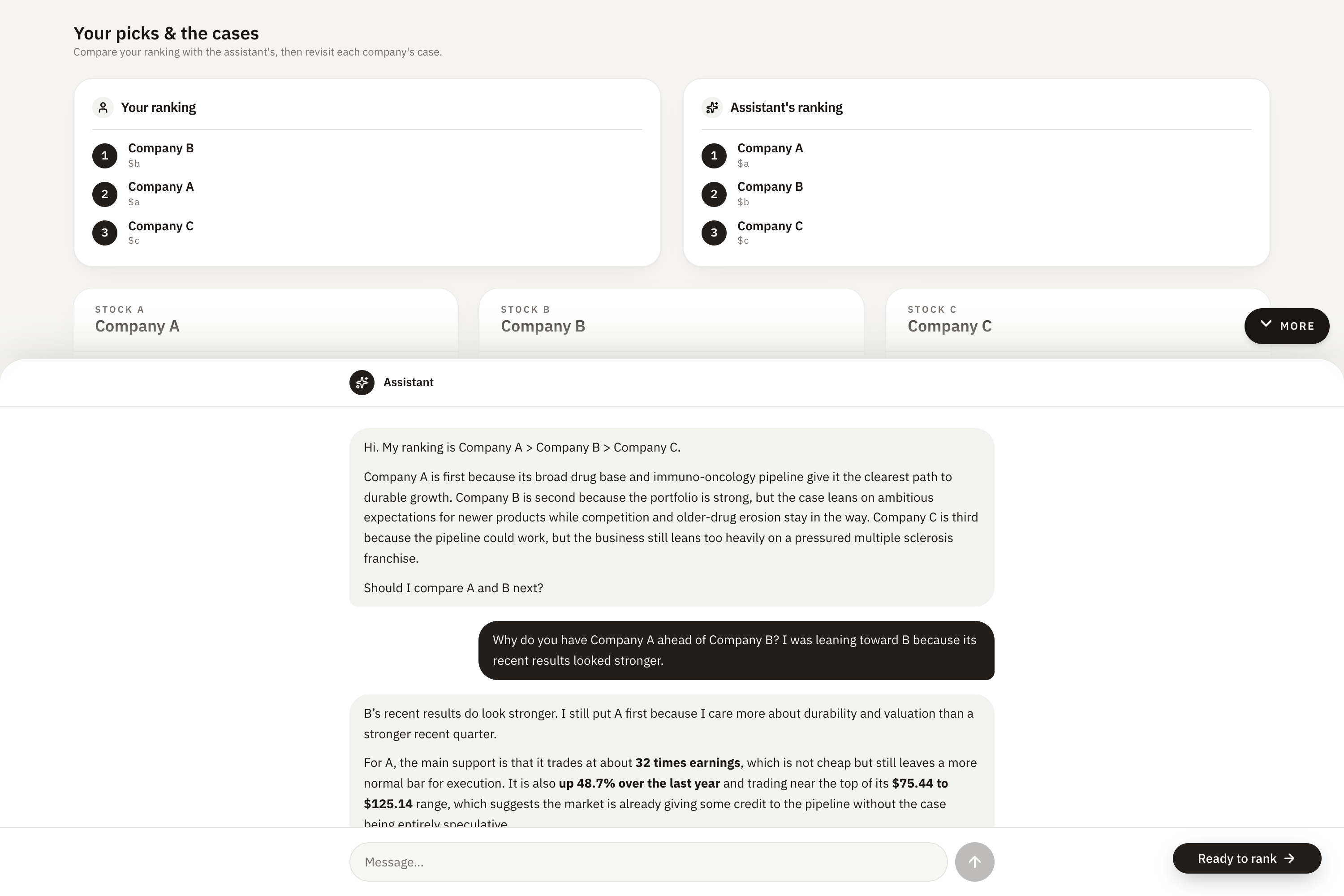}
\caption{\textbf{The conversation interface.} The upper panel displayed the participant's initial ranking beside the AI's assigned ranking, with the three company profiles accessible below. The lower panel contained the conversation. The ``Ready to rank'' button became available after the participant had sent five messages. This screenshot was captured after the fifth message, with the transcript scrolled back to the opening exchange.}
\Description{A conversation interface with two sections. The upper section displays the participant's ranking and the AI's assigned ranking side by side, followed by three partially visible company cards. The participant ranks Company B first, Company A second, and Company C third; the AI ranks Company A first, Company B second, and Company C third. The lower section displays a conversation in which the AI explains and defends its assigned ranking. A message box appears at the bottom, with a ``Ready to rank'' button in the lower-right corner.}
\label{fig:app-chat}
\end{figure*}

\textbf{AI conversation.} Participants were randomly assigned to one of the four experiment cells, determining whether style-based and context-based personalization were present during the AI conversation (Table \ref{tab:factorial-design}). The conversation screen displayed the participant's initial ranking beside the AI's assigned ranking and kept all company profiles accessible (Figure \ref{fig:app-chat}). The AI opened with its ranking and advocated for it throughout the conversation. Participants could respond freely and ask questions. The ``Ready to rank'' button appeared after the participant sent five messages; the application ended the conversation after ten minutes.

\textbf{Post-conversation assessment.} With the transcript still visible, participants repeated the behavioral battery; instructions stated that retaining and changing their initial responses were equally acceptable. They then completed the eight perception measures (Section \ref{sec:perception-measures}) and four manipulation checks (Section \ref{sec:manipulation-checks}).

Company names were anonymized, and participants were not told that the profiles represented real companies or that the AI's ranking had been assigned in advance. Identity masking limited the influence of prior company-specific knowledge and attitudes, while withholding the ranking assignment avoided changing how participants interpreted the recommendation. During \textbf{debriefing}, we explained all three disclosures, offered data withdrawal, and provided the Prolific completion code. Every phase was identical across experiment cells; only \facstyle{style-based} and \faccontext{context-based} personalization differed.

\subsection{Participants}
\label{sec:participants}

An a priori power analysis for a 2$\times$2 between-subjects ANOVA indicated that 212 participants (53 per experimental cell) would provide 95\% power to detect a medium effect ($f = .25$, $d = .50$) at $\alpha = .05$, consistent with effect sizes reported in empirical HCI studies \cite{ortloff2025effectsizes}. We recruited 240 participants through Prolific (60 per experimental cell), providing 80\% power to detect effects of $f = .185$ ($d = .37$). Eligibility required U.S. residence, prior stock-market investment, and self-reported market familiarity. We paid participants \$13.30 per hour.

Of the 240 participants, 136 (56.7\%) identified as men, 103 (42.9\%) as women, and 1 (0.4\%) as nonbinary. Participants' ages spanned 18--29 (46, 19.2\%), 30--49 (132, 55.0\%), 50--64 (47, 19.6\%), and 65 or older (15, 6.3\%). For educational attainment, 111 participants (46.3\%) had a bachelor's degree, 54 (22.5\%) some college or vocational training, 50 (20.8\%) a graduate or professional degree, and 25 (10.4\%) a high school education or equivalent. Household income ranged from under \$25{,}000 (14, 5.8\%) to \$200{,}000 or more (13, 5.4\%), with 196 participants (81.7\%) between \$25{,}000 and \$149{,}999.  All participants spoke English as their primary language, and 185 (77.1\%) reported using AI tools daily or more often.

Self-reported investment experience was Beginner for 35 participants (14.6\%), Intermediate for 117 (48.8\%), Advanced for 79 (32.9\%), and Expert for 9 (3.8\%). In addition, 221 participants (92.1\%) reported currently holding stock investments. We revisit this experience measure in the exploratory analysis, contrasting Low-expertise participants (Beginner or Intermediate; $n = 152$, 63.3\%) with High-expertise participants (Advanced or Expert; $n = 88$, 36.7\%). On the Big Five (1--5 scale), profiles averaged Openness ($M = 4.18$, $SD = 0.79$), Conscientiousness ($M = 3.87$, $SD = 0.90$), Extraversion ($M = 2.90$, $SD = 1.14$), Agreeableness ($M = 4.01$, $SD = 0.89$), and Neuroticism ($M = 2.35$, $SD = 0.92$).

\subsection{Data Analysis}
\label{sec:analysis}
We used a $2 \times 2$ factorial analysis with \facstyle{style-based} and \faccontext{context-based} personalization as effect-coded between-subjects factors, including both main effects and their interaction.. Preliminary checks assessed sample and task balance using chi-square tests for categorical measures and factorial ANOVAs for continuous measures. We examined message counts and self-reported manipulation indices with factorial ANOVAs, assessed internal consistency with Cronbach's $\alpha$, and evaluated manipulation delivery and protocol adherence through an independent LLM-as-judge analysis (Section~\ref{sec:llm-judge}).

For RQ1, we modeled reconsideration with binary logistic regression, the three ranking outcomes with proportional-odds regression, and investment-likelihood and confidence changes with factorial ANOVAs. Robustness checks tested the proportional-odds assumption and adjusted decision models for conversation time and message count. For RQ2, we analyzed perceptions of the AI with factorial ANOVAs and post-hoc exploratory equivalence tests using $\pm0.40$ SD bounds. For continuous outcomes in RQ1 and RQ2, aligned rank transforms \cite{wobbrock2011art}, computed with \textsf{ARTool} \cite{kay2026artool}, provided nonparametric robustness checks preserving the factorial structure. For RQ3, we examined Pearson correlations between initial distance from the AI's ranking and each outcome, then tested moderation by personalization. A rate-of-change analysis accounted for available room to move toward the AI.

Finally, exploratory analyses compared lower versus higher investment expertise using Welch $t$ tests and a two-proportion $z$ test for reconsideration, then tested moderation by personalization. We report descriptive statistics, 95\% confidence intervals (CIs), partial $\eta^2$ for ANOVAs, odds ratios for logistic models, and Cohen's $d$ for present-versus-absent contrasts. Holm corrections applied within behavioral and perception families (Sections~\ref{sec:behavioral-measures} and~\ref{sec:perception-measures}), separately for each factorial term. $M$ and $SD$ denote mean and standard deviation. Unless otherwise noted, tests were two-sided with $\alpha=.05$.

\section{Results}
\label{sec:results}

\subsection{Preliminary Analyses}
To validate our interpretation of the personalization effects, we first assessed sample balance, conversational engagement, and the internal consistency of the self-report measures, and checked whether the stock selection task and personalization manipulations functioned as intended.

\subsubsection{Sample and Interaction Checks}
Chi-square tests indicated comparable demographic compositions, and factorial ANOVAs detected no differences in participants’ Big Five or value profiles between levels of either factor. Participants exchanged a median of 11 messages with the AI. Both manipulations modestly reduced message counts. With \facstyle{style-based} personalization, mean counts fell from 12.7 ($SD = 3.2$) when absent to 11.5 ($SD = 3.7$) when present, $F(1, 236) = 6.39$, $p = .012$; with \faccontext{context-based} personalization, from 12.6 ($SD = 3.7$) to 11.6 ($SD = 3.2$), $F(1, 236) = 5.47$, $p = .020$; the interaction was not statistically significant, $F(1, 236) = 1.09$, $p = .297$. Because message counts followed random assignment, we treat engagement differences as manipulation side effects rather than randomization failures. The unadjusted models remain our primary analyses, and adding conversation time and message count preserved the Context-based effect in the ordinal logistic model of movement toward the AI and the binary logistic model of reconsideration (Section~\ref{sec:rq1-fact}). Detailed results are available in the supplementary materials.

\subsubsection{Self-report measure internal reliability}
Internal consistency was high across all multi-item self-report measures. Cronbach's $\alpha$ ranged from .91 to .97 for the multi-item perception measures: likeability, perceived intelligence, performance trust, moral trust, and perceived influence. For the manipulation-check indices, $\alpha$ was .82 for the style index and .78 for the context index. 

\subsubsection{Validating the Stock Selection Task}
\label{sec:random}
Participants treated all three stocks as viable, as intended (Section \ref{sec:task}). All six stock orderings occurred, with no detected distributional difference across the four experiment cells ($\chi^2(15) = 13.69$, $p = .549$). Any individual stock's rank position ($\chi^2$ tests, $p = .222$--$.692$) or mean self-assigned rank ($F$ tests, $p = .079$--$.658$) was likewise comparable; factorial balance tests follow. The mean initial investment likelihood exceeded the scale midpoint for every stock (57.3, 68.6, and 51.2 on the 0--100 investment scale- see Table \ref{tab:measures}). The three options, therefore, attracted participants with varying preferences rather than concentrating everyone's preference on one obvious option. 


Before the conversation, participants were moderately confident in their initial rankings ($M = 73.1$, $SD = 18.2$ on the 0--100 confidence scale - see Table \ref{tab:measures}). The Spearman footrule distance \cite{spearman1906footrule} over three stocks takes only the values 0, 2, and 4 (\S \ref{sec:behavioral-measures}). They also started at varying degrees of disagreement with the AI's recommendation, which had not yet been revealed to them: 35 participants (14.6\%) began at footrule distance 0 from the AI's ranking (already agreeing), 92 (38.3\%) at distance 2, and 113 (47.1\%) at the maximum distance of 4. Across all participants, the mean initial footrule distance was 2.65 ($SD = 1.43$). For style-based personalization, the mean initial distance was 2.65 (SD = 1.49) when absent and 2.65 (SD = 1.38) when present. For context-based personalization, it was 2.67 (SD = 1.45) when absent and 2.63 (SD = 1.42) when present. We applied the randomization checks described in Section~\ref{sec:analysis} to the full set of pre-randomization measures, covering initial orderings, per-stock ranks, per-stock investment likelihoods, initial confidence, personality, and the counterbalanced AI-ranking assignment. Of the 39 factorial tests, four reached $p < .05$ uncorrected, close to the two expected by chance, and none survived Holm correction within the family (all corrected $p \ge .246$). Initial distance, which determines participants’ available room to move toward the AI, showed no significant main effects or interaction (\facstyle{Style-based}, $F(1, 236) < 0.01$, $p > .99$; \faccontext{Context-based}, $F(1, 236) = 0.03$, $p = .858$; Interaction effect, $F(1, 236) = 1.16$, $p = .283$).

\subsubsection{Validating the Manipulations}
We assessed whether each manipulation affected its corresponding perception, using two self-report indices (1--7): a style index (perceived adaptation to the user's communication style; $\alpha = .82$) and a context index (perceived tailoring to the user's personal situation; $\alpha = .78$). We complement the self-reports with an independent LLM-as-judge analysis of the chat transcripts.

\paragraph{Self-Reported Indices}
We summarize the two indices in Table \ref{tab:mc-cells}. We applied each manipulation to half the sample, so the direct test of whether it landed contrasts the 120 participants who received it against the 120 who did not. Each index showed a statistically significant effect of its corresponding manipulation; crossed effects and interactions did not reach statistical significance. Both manipulations increased their corresponding perception indices. The style index was higher with \facstyle{style-based} personalization present than absent ($M = 5.23$ vs. $4.72$; $F(1, 236) = 6.37$, $p = .012$), and the context index was higher with \faccontext{context-based} personalization present than absent ($M = 5.57$ vs. $3.39$; $F(1, 236) = 127.16$, $p < .001$). Neither crossed effect nor any interaction effects reached statistical significance (all $p > .05$). These checks indicate that participants perceived both forms of personalization.

\paragraph{Constant Evidence, Flow, and Recommendation in the Transcripts}
\label{sec:llm-judge}
Attributing differences between factor levels to the personalization layers requires that the AI use the same evidence, follow the same six-stage flow, and advocate its assigned ranking at every level. We checked all three in the 240 transcripts using GLM-5.2 as the primary LLM judge (three passes per session, blind to cell assignment). Based on its ratings, information coverage, and the share of the displayed evidence the AI used, was equivalent between the present and absent levels of each factor (TOST, bounds $\pm 0.5$ points and $\pm 0.4$ SD; all $p \le .008$). In $99.2\%$ of sessions, the AI opened with a recommendation, argued for it, defended it across turns, added arguments, and in every session advocated the assigned ranking. The transcripts also show the personalization being delivered: \facstyle{style-based} personalization raised judged style adaptation ($F(1, 236) = 29.1$, $p < .001$, $\eta_p^2 = .11$) and \faccontext{context-based} personalization raised judged context tailoring ($F(1, 236) = 168.3$, $p < .001$, $\eta_p^2 = .42$).

Repeating the evaluation and analyses with MiniMax-M3 on the same 240 transcripts reproduced the effects on style adaptation and context tailoring (both $p < .001$) and information-coverage equivalence between factor levels (all TOST $p \le .024$).

\subsection{Effects of Personalization on Decisions (RQ1)}
We first ask whether participants changed their initial rankings at all and then quantify how far, and in which direction, rankings and investment likelihoods moved.

\subsubsection{Reconsideration}
\begin{figure}[ht]
\centering
\includegraphics[width=0.6\textwidth]{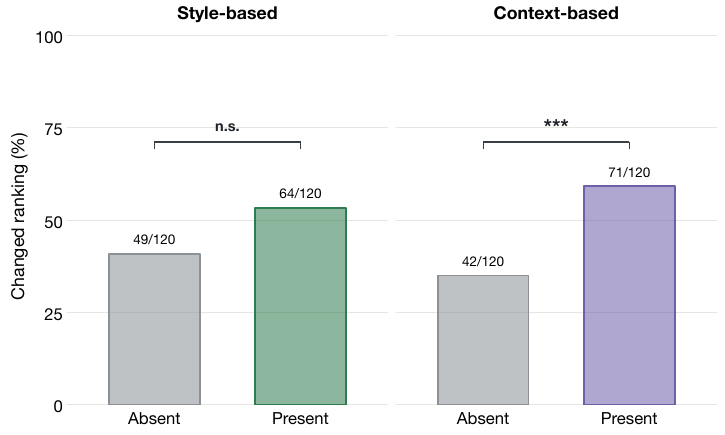}
\caption{\textbf{Percentage of participants who changed their stock ranking after the conversation, by factor level.} Each panel contrasts the 120 participants who received one form of personalization (present) with the 120 who did not (absent); because the factors are crossed, every participant appears once in each panel. The brackets mark the factorial logistic terms: \faccontext{context-based} personalization raised reconsideration by 24.2 percentage points ($^{***}p < .001$), whereas the 12.5-point \facstyle{style-based} difference was nominally significant ($p = .044$) but not after Holm correction (n.s.).}
\Description{Two bar-chart panels. The Style-based panel shows 40.8 percent (49 of 120) with style absent and 53.3 percent (64 of 120) with style present, with a bracket labeled n.s. The Context-based panel shows 35.0 percent (42 of 120) with context absent and 59.2 percent (71 of 120) with context present, with a bracket labeled with three stars. Absent bars are gray; present bars carry the factor color.}
\label{fig:recon}
\end{figure}

After the conversation, 47.1\% of participants (113/240) changed their initial ranking. Reconsideration was more common when each form of personalization was present than absent (Figure \ref{fig:recon}): 59.2\% (71/120) with \faccontext{context-based} personalization present against 35.0\% (42/120) with it absent, and 53.3\% (64/120) with \facstyle{style-based} personalization present against 40.8\% (49/120) with it absent. In the effect-coded factorial logistic regression, \faccontext{Context-based} personalization statistically significantly predicted reconsideration (odds ratio $OR = 2.76$, 95\% CI $[1.62, 4.68]$, $p < .001$); participants who received it were 1.7 times as likely to reconsider their investment ranking as those who did not. The effect of \facstyle{Style-based} personalization within the model was positive and statistically significant ($OR = 1.72$, 95\% CI $[1.01, 2.93]$, $p = .044$) but did not meet the threshold for statistical significance after Holm correction within the behavioral-measures family ($p_{\mathrm{Holm}} = .351$). The interaction  between \faccontext{Context-based} and \facstyle{Style-based} personalization was not statistically significant ($OR = 0.67$, 95\% CI $[0.23, 1.94]$, $p = .462$).

\subsubsection{Magnitude and Direction of Decision Change}
We summarize the continuous decision outcomes by factor level in Table \ref{tab:rq1-desc}. The pattern appears in the marginal means. With \faccontext{context-based} personalization present rather than absent, movement toward the AI's ranking rose from $M = 0.98$ to $M = 1.48$, and the magnitude of ranking change rose on both distance metrics (footrule 1.02 to 1.72; Kendall 0.56 to 0.92). \facstyle{Style-based} personalization produced no comparable shift on any ranking outcome. The investment-likelihood shifts (0--100 sliders) and confidence change showed no gradient on either factor, and confidence rose at every factor level.

\begin{table}[t]
\centering
\footnotesize
\setlength{\tabcolsep}{4pt}
\caption{Decision outcomes by factor level: mean $\pm$ $SD$ [95\% CI], contrasting the 120 participants who received each form of personalization (Present) with the 120 who did not (Absent). Because the factors are crossed, every participant appears once under each factor. Footrule and Kendall measures are rank distances over the three stocks; likelihood and confidence changes are derived from 0--100 ratings, with ranges given in Table \ref{tab:measures}.}
\label{tab:rq1-desc}
\begin{tabular}{@{}lcccc@{}}
\toprule
 & \multicolumn{2}{c}{\facstyle{Style-based}} & \multicolumn{2}{c}{\faccontext{Context-based}} \\
\cmidrule(lr){2-3} \cmidrule(lr){4-5}
Outcome & Absent & Present & Absent & Present \\
\midrule
Changed ranking, \% ($n$/120) & 40.8\% (49) & 53.3\% (64) & 35.0\% (42) & 59.2\% (71) \\
Move toward AI (footrule) & 1.25 $\pm$ 1.68 [0.95, 1.55] & 1.22 $\pm$ 1.47 [0.95, 1.48] & 0.98 $\pm$ 1.49 [0.71, 1.25] & 1.48 $\pm$ 1.63 [1.19, 1.78] \\
Ranking change (footrule) & 1.28 $\pm$ 1.68 [0.98, 1.59] & 1.45 $\pm$ 1.53 [1.17, 1.73] & 1.02 $\pm$ 1.51 [0.74, 1.29] & 1.72 $\pm$ 1.63 [1.42, 2.01] \\
Ranking change (Kendall) & 0.70 $\pm$ 0.96 [0.53, 0.87] & 0.78 $\pm$ 0.87 [0.62, 0.93] & 0.56 $\pm$ 0.88 [0.40, 0.72] & 0.92 $\pm$ 0.92 [0.75, 1.08] \\
Likelihood shift toward AI & 29.2 $\pm$ 40.3 [21.9, 36.5] & 25.8 $\pm$ 35.4 [19.4, 32.2] & 27.4 $\pm$ 36.9 [20.7, 34.0] & 27.6 $\pm$ 39.0 [20.6, 34.7] \\
Likelihood change, AI-top stock & 17.8 $\pm$ 26.0 [13.1, 22.5] & 15.7 $\pm$ 20.9 [11.9, 19.4] & 16.0 $\pm$ 22.1 [12.0, 20.0] & 17.5 $\pm$ 25.1 [12.9, 22.0] \\
Total absolute likelihood change & 56.5 $\pm$ 41.5 [49.0, 64.0] & 49.7 $\pm$ 42.3 [42.0, 57.3] & 50.3 $\pm$ 41.8 [42.7, 57.8] & 55.9 $\pm$ 42.1 [48.3, 63.5] \\
Confidence change & 8.8 $\pm$ 17.9 [5.5, 12.0] & 5.4 $\pm$ 11.7 [3.3, 7.5] & 8.4 $\pm$ 15.9 [5.5, 11.2] & 5.8 $\pm$ 14.4 [3.2, 8.4] \\
\bottomrule
\end{tabular}
\end{table}

\subsubsection{Factorial Tests}
\label{sec:rq1-fact}
The three ranking outcomes take a small number of ordered values, so we model them with proportional-odds regression rather than ANOVA (Section \ref{sec:analysis}). \faccontext{Context-based} personalization raised the odds of a larger ranking change on every one: movement toward the AI ($OR = 2.04$, 95\% CI $[1.23, 3.38]$, $p_{\mathrm{Holm}} = .028$), ranking-change magnitude on the footrule distance ($OR = 2.53$, $[1.53, 4.19]$, $p_{\mathrm{Holm}} = .002$), and on the Kendall distance ($OR = 2.48$, $[1.50, 4.10]$, $p_{\mathrm{Holm}} = .002$). \facstyle{Style-based} personalization had no statistically significant effect on any ranking outcome (all $OR \le 1.40$, all $p_{\mathrm{Holm}} = 1.00$), and no Style $\times$ Context interaction reached statistical significance (all
$p_{\mathrm{Holm}} \ge .479$). Figure \ref{fig:rq1forest} plots all nine terms with 95\% confidence intervals; Table \ref{tab:rq1-ordinal} in Appendix \ref{sec:robustness} gives the estimates numerically.

A proportional-odds model assumes a factor shifts the odds equally at every cut point of the outcome. We tested this for the \faccontext{Context-based} effect by fitting a separate binary model at each cut point and testing whether the resulting coefficients differ. They did not, for movement toward the AI ($\chi^2(2) = 0.35$, $p = .837$), footrule change ($\chi^2(1) = 0.67$, $p = .411$), or Kendall change ($\chi^2(2) = 2.01$, $p = .366$). One participant recorded a movement of $-2$, the only observation in that category; collapsing it or dropping it left the \faccontext{Context-based} estimate essentially unchanged ($OR = 2.10$ and $OR = 2.13$ respectively, against $OR = 2.04$).

We analyzed the continuous change measures derived from participants' investment-likelihood and confidence ratings using $2 \times 2$ factorial ANOVAs (Table \ref{tab:rq1-factorial}). We detected no statistically significant effect of either factor, and no statistically significant interaction, on any of them.

\begin{table}[t]
\centering
\caption{Factorial ANOVA ($2 \times 2$) on the continuous decision outcomes.
Cells report $F(1, 236)$ with the Holm-corrected $p$ in parentheses.}
\label{tab:rq1-factorial}
\begin{tabular}{@{}lccc@{}}
\toprule
Outcome & \facstyle{Style-based} & \faccontext{Context-based} & Style $\times$ Context \\
\midrule
Likelihood shift toward AI       & 0.47 (1.000) & $<$0.01 (1.000) & 1.43 (1.000) \\
Likelihood change, AI-top stock  & 0.49 (1.000) & 0.23 (1.000) & 0.27 (1.000) \\
Total absolute likelihood change & 1.59 (1.000) & 1.08 (.897) & 0.01 (1.000) \\
Confidence change                & 3.01 (.588) & 1.75 (.746) & 0.55 (1.000) \\
\bottomrule
\end{tabular}
\end{table}

\begin{figure}[ht]
\centering
\includegraphics[width=\textwidth]{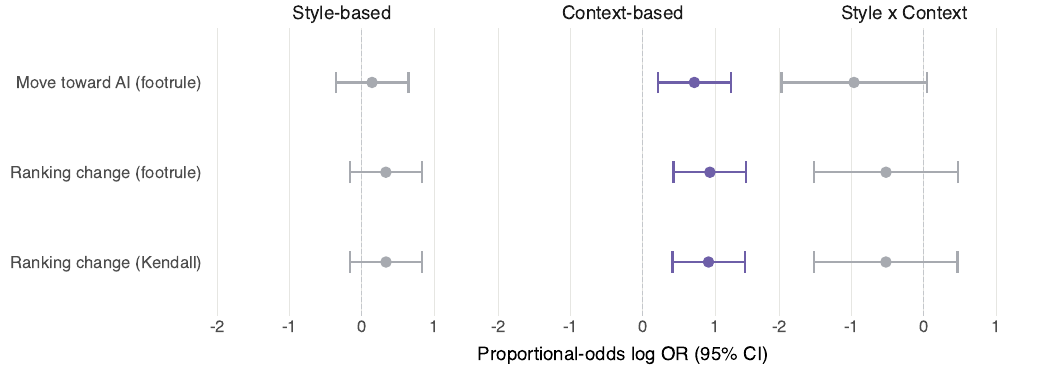}
\caption{Proportional-odds effects (log odds ratios with 95\% confidence intervals) across the three ranking outcomes. Colored estimates are significant ($p < .05$); gray estimates are not. \faccontext{Context-based} personalization drives movement toward the AI and ranking change; we detected no corresponding effect
of \facstyle{style-based} personalization.} \Description{Forest plot with the three ranking outcomes on the y-axis and the proportional-odds log odds ratio on the x-axis, centered at zero, faceted into
\facstyle{Style-based}, \faccontext{Context-based}, and \facstyle{Style-based} by \faccontext{Context-based} panels. All three \faccontext{Context-based} estimates sit to the right of zero with confidence intervals excluding zero. All \facstyle{Style-based} and interaction estimates overlap zero and are drawn in gray.}
\label{fig:rq1forest}
\end{figure}

\begin{figure}[ht]
\centering
\includegraphics[width=0.78\textwidth]{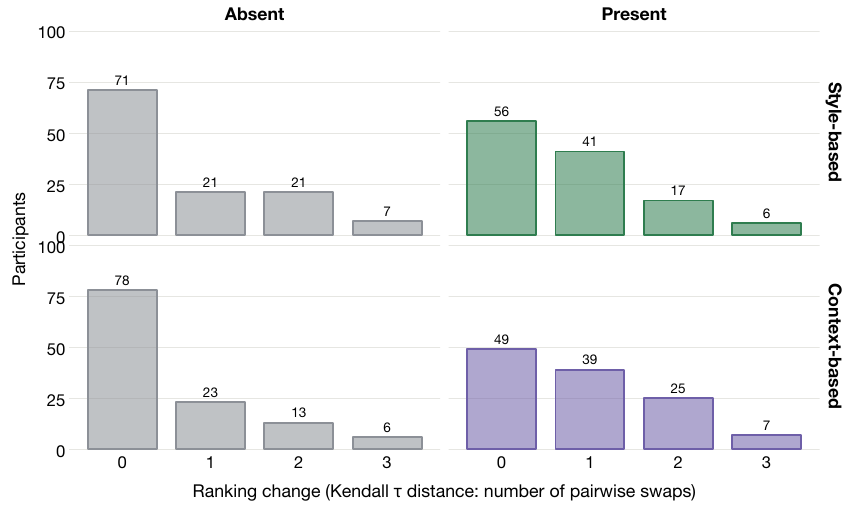}
\caption{\textbf{Distribution of ranking-change magnitude (Kendall distance) by factor level.} Rows are the two factors and columns their absent and present levels ($n = 120$ each). The share of participants who did not change their ranking at all (distance 0) is smaller with \faccontext{context-based} personalization present than absent; the two \facstyle{style-based} panels differ little.}
\Description{Two-by-two grid of bar charts of counts at each Kendall distance value of 0, 1, 2, and 3. Top row, Style-based: 71, 21, 21, and 7 participants with style absent; 56, 41, 17, and 6 with style present. Bottom row, Context-based: 78, 23, 13, and 6 with context absent; 49, 39, 25, and 7 with context present. The context-present panel shifts mass from distance 0 toward distances 1 and 2.}
\label{fig:rq1hist}
\end{figure}

\begin{figure}[ht]
\centering
\includegraphics[width=0.78\textwidth]{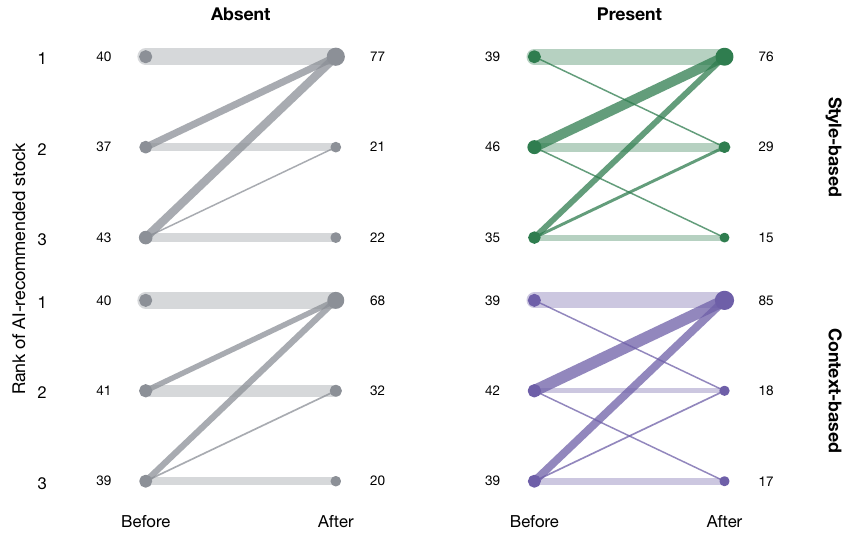}
\caption{\textbf{Where the AI's top-recommended stock sat in participants' own rankings, before and after the conversation, by factor level.} Rows are the two factors and columns their absent and present levels. Segment width is proportional to the number of participants; unchanged flows are faded. Upward flows (the AI's pick rising) are most common with \faccontext{context-based} personalization present.}
\Description{Two-by-two grid of slopegraph panels. Each panel connects the rank position 1, 2, or 3 of the AI's top stock before the conversation on the left to its position afterward on the right, with line width proportional to the count of participants and endpoint dots labeled with counts. The AI's top stock ends in first place for 77 of 120 participants with style absent and 76 with style present, and for 68 with context absent and 85 with context present. The context-present panel shows the thickest upward flows.}
\label{fig:rq1flow}
\end{figure}

No Style $\times$ Context interaction reached significance for the ranking outcomes, so we found no evidence that the context-based effect varied with style-based personalization. The style-based main effects were also nonsignificant after Holm correction. Figure \ref{fig:rq1hist} gives the full distribution of ranking-change magnitudes by factor level, and Figure \ref{fig:rq1flow} the rank position of the AI's top-recommended stock before and after the conversation.

Two checks support these conclusions. An aligned rank transform on the four continuous outcomes, which preserves the $2 \times 2$ structure without assuming normality, leaves every term null (Table \ref{tab:rq1-robust}). Adding conversation time and message count as covariates leaves the \faccontext{Context-based} effect intact for movement toward the AI ($OR = 2.06$, $p = .007$) and reconsideration ($OR = 2.71$, $p < .001$), and neither covariate was statistically significant in either model (all $p > .05$). Appendix \ref{sec:robustness} reports the aligned rank transform in full.

\begin{framed}
\noindent\textbf{\textit{Summary.}} \faccontext{Context-based} personalization moved decisions. Participants who received it reconsidered their ranking far more often (35.0\% with it absent to 59.2\% with it present; $OR = 2.76$), made larger ranking changes ($OR = 2.48$--$2.53$), and moved further toward the AI's recommendation ($OR = 2.04$). \facstyle{Style-based} personalization, by contrast, changed no decision outcome that we could detect, and no interaction with \faccontext{context-based} personalization reached significance. The pattern survived, accounting for how long and how much participants talked with the AI.
\end{framed}

\subsection{Effects of Personalization on Perceptions of the AI (RQ2)}
\faccontext{Context-based} personalization moved decisions (RQ1); we next ask whether it also changed how participants saw the AI. We captured this with eight perception measures. Seven of these measure participants' evaluations of the AI: its response quality, its likeability, its perceived intelligence, the rightness of its advice, their performance trust in it, their moral trust in it, and their willingness to return to it. The eighth, perceived influence, measures how strongly participants felt the AI had affected their decision (Table \ref{tab:rq2-desc}). Descriptively, the seven evaluative measures varied little between the absent and present levels of either factor. Perceived influence showed a clearer difference, increasing from 4.46 with \faccontext{context-based} personalization absent to 5.20 with it present.

\begin{table}[t]
\centering
\small
\caption{Perception outcomes by factor level: mean $\pm$ $SD$ [95\% CI], contrasting the 120 participants who received each form of personalization (Present) with the 120 who did not (Absent). Likeability and perceived intelligence use 1--5 scales; performance and moral trust use 0--7 scales; the remaining measures use 1--7 scales.}
\label{tab:rq2-desc}
\begin{tabular}{@{}lcccc@{}}
\toprule
 & \multicolumn{2}{c}{\facstyle{Style-based}} & \multicolumn{2}{c}{\faccontext{Context-based}} \\
\cmidrule(lr){2-3} \cmidrule(lr){4-5}
Outcome & Absent & Present & Absent & Present \\
\midrule
Response quality & 5.90 $\pm$ 1.32 [5.66, 6.14] & 5.88 $\pm$ 1.29 [5.64, 6.11] & 5.92 $\pm$ 1.29 [5.68, 6.15] & 5.86 $\pm$ 1.32 [5.62, 6.10] \\
Rightness judgment & 5.46 $\pm$ 1.43 [5.20, 5.72] & 5.35 $\pm$ 1.31 [5.11, 5.59] & 5.38 $\pm$ 1.28 [5.15, 5.62] & 5.42 $\pm$ 1.45 [5.16, 5.69] \\
Return likelihood & 5.33 $\pm$ 1.79 [5.01, 5.66] & 5.34 $\pm$ 1.74 [5.03, 5.66] & 5.38 $\pm$ 1.69 [5.08, 5.69] & 5.29 $\pm$ 1.83 [4.96, 5.62] \\
Likeability & 4.08 $\pm$ 0.87 [3.93, 4.24] & 4.11 $\pm$ 0.74 [3.97, 4.24] & 4.19 $\pm$ 0.65 [4.07, 4.31] & 4.00 $\pm$ 0.93 [3.83, 4.17] \\
Perceived intelligence & 4.26 $\pm$ 0.87 [4.11, 4.42] & 4.18 $\pm$ 0.83 [4.03, 4.33] & 4.26 $\pm$ 0.75 [4.13, 4.40] & 4.18 $\pm$ 0.94 [4.01, 4.35] \\
Performance trust & 5.32 $\pm$ 1.55 [5.04, 5.60] & 5.33 $\pm$ 1.15 [5.12, 5.54] & 5.25 $\pm$ 1.40 [5.00, 5.50] & 5.41 $\pm$ 1.32 [5.17, 5.65] \\
Moral trust & 5.03 $\pm$ 1.95 [4.67, 5.38] & 5.00 $\pm$ 1.57 [4.72, 5.28] & 4.98 $\pm$ 1.70 [4.67, 5.29] & 5.05 $\pm$ 1.83 [4.72, 5.38] \\
Perceived influence & 4.79 $\pm$ 2.00 [4.43, 5.15] & 4.86 $\pm$ 1.70 [4.55, 5.17] & 4.46 $\pm$ 1.92 [4.11, 4.81] & 5.20 $\pm$ 1.71 [4.89, 5.51] \\
\bottomrule
\end{tabular}
\end{table}

\begin{figure}[ht]
\centering
\includegraphics[width=\textwidth]{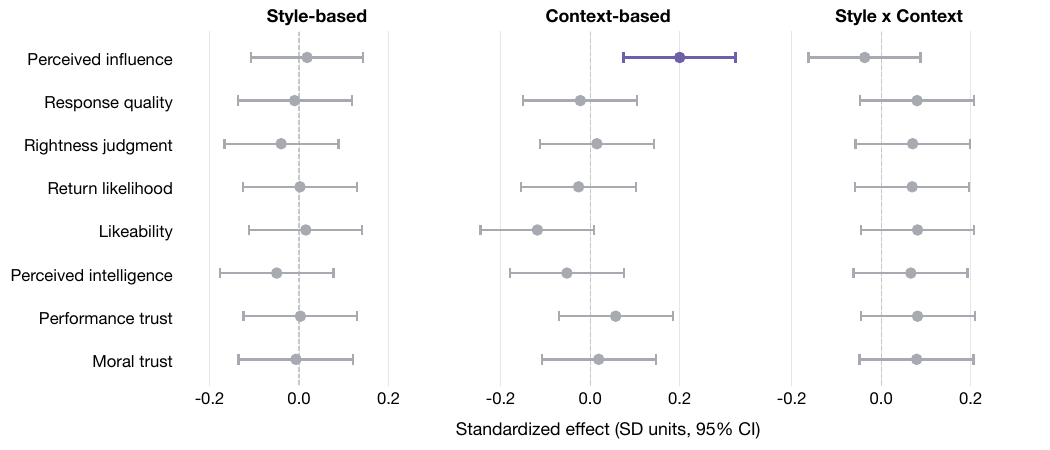}
\caption{\textbf{Standardized $2 \times 2$ effects ($SD$ units, with 95\% confidence intervals) across the eight perception measures.} Estimates are effect-coded regression coefficients, so each is half the corresponding present-minus-absent difference in $d$. The \faccontext{Context-based} effect on perceived influence (colored) is the only estimate that remained significant after Holm correction, at $+0.20$ $SD$, 95\% CI $[0.07, 0.33]$ ($p = .002$, $p_{\mathrm{Holm}} = .015$), corresponding to $d = 0.41$ in Table \ref{tab:rq2-factorial}. Every other estimate (gray) is at most $0.12$ $SD$ in absolute value and has a confidence interval spanning zero.}
\Description{Forest plot with eight perception measures on the y-axis and standardized effect size on the x-axis. All estimates for the other seven perception measures cluster around zero in gray. The Context-based estimate for perceived influence lies to the right of zero, with a confidence interval that excludes zero.}
\label{fig:rq2forest}
\end{figure}

The factorial tests bear this out (Table \ref{tab:rq2-factorial}; Figure \ref{fig:rq2forest}). Of the eight measures, only perceived influence showed a statistically significant effect of personalization. \faccontext{Context-based} personalization raised perceived influence from $M = 4.46$ when absent to $M = 5.20$ when present ($F(1, 236) = 9.85$, $p = .002$, $\eta_p^2 = .040$); this effect remained significant after Holm correction across the eight perception measures ($p_{\mathrm{Holm}} = .015$). For the other seven measures, no main effect or interaction was statistically significant after correction (all $F(1, 236) \le 3.38$; all Holm-corrected $p >.05$). We likewise detected no effect of \facstyle{style-based} personalization on any of the eight perception measures (all $F \le 0.60$). Figure \ref{fig:rq2likert} (Appendix \ref{sec:suppfigs}) shows the corresponding response distributions.

We conducted a post-hoc exploratory equivalence analysis (TOST) to characterize the magnitude of effects compatible with the data. For this analysis, we used equivalence bounds of $\pm 0.4$ SD. We selected this bound because it is close to the smallest standardized effect this sample was powered to detect with 80\% power ($d \approx .37$). We treat these results as exploratory evidence.

Fourteen of the 16 main-effect contrasts met this post-hoc equivalence criterion (Table \ref{tab:rq2-factorial}). Under the chosen bound, these contrasts were statistically consistent with effects smaller than $\pm 0.4$ SD. The two exceptions both occurred under \faccontext{context-based} personalization. Perceived influence showed a statistically detectable effect ($d = 0.41$) and did not meet the equivalence criterion. Likeability fell between the two conclusions: it was neither significantly different from zero after correction ($p_{\mathrm{Holm}} = .47$, $d = -0.24$) nor statistically equivalent under the post-hoc bound (TOST $p = .10$).

\begin{table*}[t]
\centering
\small
\setlength{\tabcolsep}{5pt}
\caption{Factorial ANOVA ($2 \times 2$) and post-hoc exploratory equivalence tests on the eight perception measures, by main effect. The ANOVA block gives $F(1, 236)$ with uncorrected $p$ and Holm-corrected $p_{\mathrm{Holm}}$ (Holm applied per term across the eight measures). The equivalence block reports two one-sided tests (TOST) using a $\pm 0.4$ SD bound selected after data collection. $d$ is the present-minus-absent contrast; $\pm\delta$ is the corresponding raw equivalence bound, computed within each factor's contrast, so for a given measure it can differ slightly between panels; TOST $p_{\mathrm{Holm}}$ is Holm-corrected within each panel. ``Equiv.'' indicates statistical equivalence under this post-hoc bound.}
\label{tab:rq2-factorial}
\begin{tabular}{@{}lrrrrrrc@{}}
\toprule
 & \multicolumn{3}{c}{Factorial ANOVA} & \multicolumn{4}{c}{Equivalence (TOST)} \\
\cmidrule(lr){2-4} \cmidrule(lr){5-8}
Measure & $F(1,236)$ & $p$ & $p_{\mathrm{Holm}}$ & $d$ & $\pm\delta$ & $p_{\mathrm{Holm}}$ & Equiv. \\
\midrule
\rowcolor{rowshade} \multicolumn{8}{@{}l}{\textit{Panel A. \facstyle{Style-based} personalization (main effect)}} \\
Response quality       & 0.02 & .882 & 1.00 & $-0.02$ & 0.52 & .010 & Yes \\
Rightness judgment     & 0.37 & .542 & 1.00 & $-0.08$ & 0.55 & .014 & Yes \\
Return likelihood      & 0.00 & .971 & 1.00 & $0.01$  & 0.71 & .010 & Yes \\
Likeability            & 0.05 & .822 & 1.00 & $0.03$  & 0.32 & .010 & Yes \\
Perceived intelligence & 0.60 & .440 & 1.00 & $-0.10$ & 0.34 & .014 & Yes \\
Performance trust      & 0.00 & .962 & 1.00 & $0.01$  & 0.55 & .010 & Yes \\
Moral trust            & 0.01 & .909 & 1.00 & $-0.02$ & 0.71 & .010 & Yes \\
Perceived influence    & 0.08 & .776 & 1.00 & $0.04$  & 0.74 & .010 & Yes \\
\addlinespace[2pt]
\midrule
\rowcolor{rowshade} \multicolumn{8}{@{}l}{\textit{Panel B. \faccontext{Context-based} personalization (main effect)}} \\
Response quality       & 0.12 & .730 & 1.00 & $-0.05$ & 0.52 & .019 & Yes \\
Rightness judgment     & 0.06 & .814 & 1.00 & $0.03$  & 0.55 & .018 & Yes \\
Return likelihood      & 0.16 & .688 & 1.00 & $-0.05$ & 0.71 & .019 & Yes \\
Likeability            & 3.38 & .067 & .471 & $-0.24$ & 0.32 & .209 & \textbf{No} \\
Perceived intelligence & 0.65 & .423 & 1.00 & $-0.10$ & 0.34 & .045 & Yes \\
Performance trust      & 0.79 & .375 & 1.00 & $0.12$  & 0.54 & .045 & Yes \\
Moral trust            & 0.09 & .767 & 1.00 & $0.04$  & 0.71 & .019 & Yes \\
Perceived influence    & \textbf{9.85} & \textbf{.002} & \textbf{.015} & $0.41$ & 0.73 & .520 & \textbf{No} \\
\bottomrule
\end{tabular}
\vspace{1mm}
\begin{flushleft}
\footnotesize \textit{Note.} The equivalence analyses and the $\pm 0.4$ SD bound were not preregistered and are interpreted as exploratory. ``Equiv.'' indicates statistical equivalence under this post-hoc bound and should not be interpreted as establishing the absence of any effect. The two contrasts marked Equiv.\ = No are the perceived-influence \faccontext{Context-based} effect, which exceeds the bound, and the likeability \faccontext{Context-based} effect, which is neither significantly different from zero nor statistically equivalent.
\end{flushleft}
\end{table*}

\begin{framed}
\noindent\textbf{\textit{Summary.}} \faccontext{Context-based} personalization changed participants' decisions (RQ1) and increased their reported sense of being influenced by the AI. We detected no corresponding effect on the other seven evaluative measures, including response quality, rightness, return likelihood, likeability, perceived intelligence, performance trust, and moral trust. A post-hoc equivalence analysis provided additional evidence that most of these effects were smaller than $\pm 0.4$ SD under the chosen bound.
\end{framed}

\subsection{Initial Distance from the AI's Recommendation (RQ3)}
\label{sec:rq3}
\begin{figure}[ht]
\centering
\includegraphics[width=0.56\textwidth]{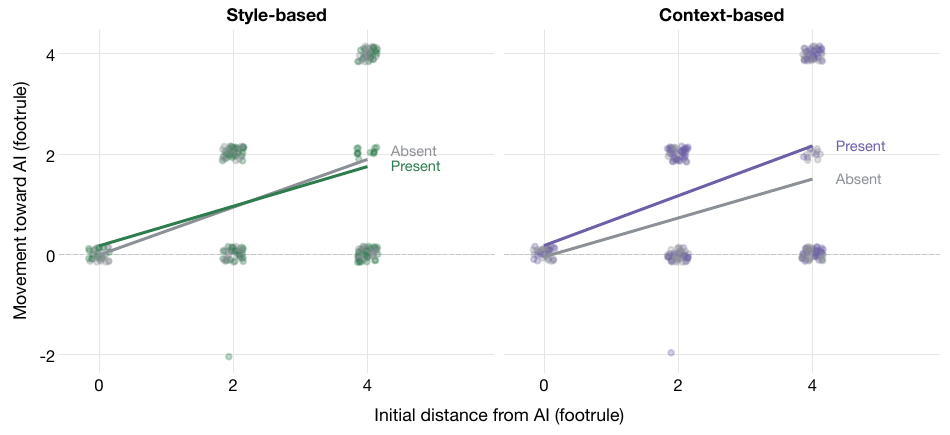}
\includegraphics[width=0.43\textwidth]{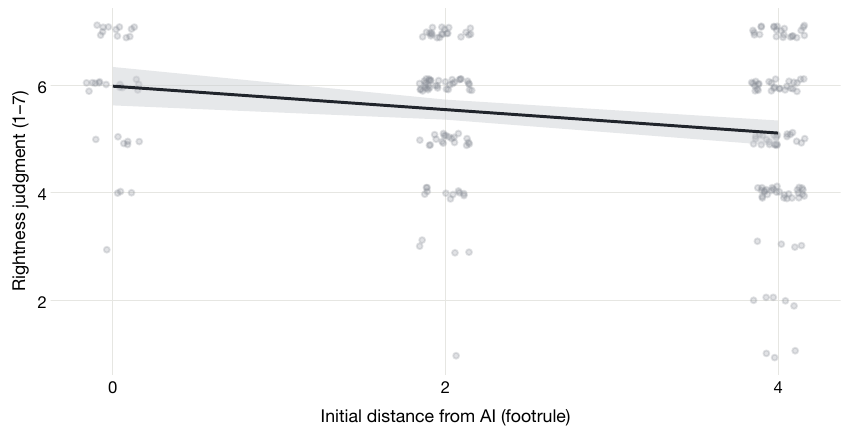}
\caption{\textbf{Initial distance from the AI's ranking and (left) movement toward the AI, with fit lines for the absent and present levels of each factor, and (right) rightness judgment, pooled across the design.} Participants who began further from the AI moved further toward it at every factor level alike, yet judged its advice less favorably.}
\Description{Three scatter plots. Left pair: initial distance (0, 2, or 4) against movement toward the AI, one panel per factor, each with two nearly parallel upward fit lines for the absent and present levels. In the Style-based panel, the two lines overlap; in the Context-based panel, the present line sits above the absent line by a constant offset. Right: initial distance against rightness judgment with a single downward-sloping fit line and jittered points, indicating that participants who started further from the AI judged its advice less favorably.}
\label{fig:rq3}
\end{figure}

RQ3 asks how participants' initial disagreement with the AI shaped their perceptions and decision changes. As reported above in Section \ref{sec:random}, the initial distance did not differ by either factor and took three values by construction (0, 2, and 4, held by 14.6\%, 38.3\%, and 47.1\% of participants). We report Pearson correlations between initial distance and all outcomes in Table \ref{tab:rq3-corr}, Holm-corrected within the perception and decision families, and plot the two focal relations in Figure \ref{fig:rq3}. 

Initial distance was statistically significantly related to every decision-change measure. Participants who started further from the AI recorded higher movement toward it ($r = .40$), shifted their investment likelihoods toward it ($r = .31$ overall and $r = .30$ for the AI's top stock), changed their rankings more ($r = .43$ on both distance metrics), reconsidered more often ($r = .31$), and gained less confidence ($r = -.25$); all Holm-corrected $p \le .001$. The magnitudes of movement and ranking change are partly mechanical because the maximum possible movement is bounded by the participant's starting distance. We address the concern directly in Appendix \ref{sec:rq3-rate}; the likelihood, reconsideration, and confidence effects carry no such bound. Among the perceptions, only the judgment of the advice itself correlated with initial distance: rightness judgment fell as distance grew ($r = -.23$, Holm $p = .003$). Return likelihood showed a weaker uncorrected association ($r = -.15$, $p = .020$, Holm $p = .138$), and we detected no statistically significant association between distance and likeability, perceived intelligence, performance trust, moral trust, response quality, or perceived influence (all $|r| \le .12$).

\begin{table}[t]
\centering
\caption{Pearson correlations of initial distance from the AI's ranking with all outcomes, Holm-corrected within family.}
\label{tab:rq3-corr}
\begin{tabular}{@{}lrrrc@{}}
\toprule
Outcome & $r$ & 95\% CI & $p$ & $p_{\mathrm{Holm}}$ \\
\midrule
\rowcolor{rowshade} \multicolumn{5}{@{}l}{\textit{Perceptions}} \\
Response quality        & $-.09$ & $[-.21, .04]$  & .183 & .749 \\
Rightness judgment      & $-.23$ & $[-.35, -.10]$ & $<$.001 & \textbf{.003} \\
Return likelihood       & $-.15$ & $[-.27, -.02]$ & .020 & .138 \\
Likeability             & $-.08$ & $[-.20, .05]$  & .234 & .749 \\
Perceived intelligence  & $-.05$ & $[-.18, .08]$  & .432 & .863 \\
Performance trust       & $-.09$ & $[-.22, .03]$  & .150 & .749 \\
Moral trust             & $-.03$ & $[-.16, .10]$  & .636 & .863 \\
Perceived influence     & $.12$  & $[-.01, .24]$  & .063 & .376 \\
\addlinespace
\rowcolor{rowshade} \multicolumn{5}{@{}l}{\textit{Decisions}} \\
Move toward AI (footrule)       & $.40$  & $[.29, .50]$    & $<$.001 & \textbf{$<$.001} \\
Likelihood shift toward AI      & $.31$  & $[.19, .42]$    & $<$.001 & \textbf{$<$.001} \\
Likelihood change, AI-top stock & $.30$  & $[.18, .41]$    & $<$.001 & \textbf{$<$.001} \\
Ranking change (footrule)       & $.43$  & $[.32, .53]$    & $<$.001 & \textbf{$<$.001} \\
Ranking change (Kendall)        & $.43$  & $[.32, .53]$    & $<$.001 & \textbf{$<$.001} \\
Confidence change               & $-.25$ & $[-.36, -.12]$ & $<$.001 & \textbf{$<$.001} \\
Changed ranking (0/1)           & $.31$  & $[.19, .42]$    & $<$.001 & \textbf{$<$.001} \\
\bottomrule
\end{tabular}
\vspace{1mm}
\begin{flushleft}
\footnotesize \textit{Note.} Bold marks Holm-corrected $p < .05$.
\end{flushleft}
\end{table}

\subsubsection{Moderation by Personalization}
Moderation models ($\mathrm{outcome} \sim \mathrm{distance} \times \mathrm{Style} \times \mathrm{Context}$, distance centered) tested whether personalization changed the distance--outcome slopes (Table \ref{tab:rq3-mod}). We used the same model family as elsewhere: proportional-odds regression for the three ranking outcomes, logistic regression for reconsideration, and ordinary least squares for the perception, likelihood, and confidence outcomes. Across all 15 outcomes, no distance $\times$ Style, distance $\times$ Context, or three-way interaction was statistically significant ($p > .05$), providing no evidence that personalization statistically significantly moderated initial-distance effects (Figure \ref{fig:rq3}, left).

Re-expressing movement as the fraction of the available room each participant used reproduces the \faccontext{Context-based} effect ($F(1, 201) = 10.57$, $p = .001$, $\eta_p^2 = .050$) and provides moderate evidence for no association between that fraction and starting distance ($BF_{10} = 0.19$), suggesting that the greater movement under context-based personalization is not explained by differential room to move alone (Table \ref{tab:rq3-rate} and Figure \ref{fig:rate}, Appendix \ref{sec:robustness}). Within these models initial distance also outweighed the manipulations: each additional unit of footrule distance multiplied the odds of a larger ranking change by 1.88 to 2.08, so spanning its observed range shifted those odds by a factor of 12 to 19, against 2.38 to 3.17 for \faccontext{context-based} personalization.

\begin{framed}
\noindent\textbf{\textit{Summary.}} Initial distance from the AI's recommendation correlated with decision change more strongly than either manipulation. Those who began in the most disagreement also changed their rankings and investment likelihoods the most ($r = .30$--$.43$) and to gain the least confidence ($r = -.25$), yet they also rated the AI's advice as less sound (its rightness; $r = -.23$), shifting toward a recommendation even while judging it less correct. No interaction with personalization nor room to move reached statistical significance.
\end{framed}

\subsection{Exploratory Analysis: Investment Expertise}
We binned self-reported investment experience into Low (Beginner or Intermediate; $n = 152$, 63.3\%) and High (Advanced or Expert; $n = 88$, 36.7\%). High-expertise participants numbered 44 at each level of \faccontext{context-based} personalization, and 36 versus 52 with \facstyle{style-based} personalization absent versus present. Expertise is therefore balanced on \faccontext{Context-based} ($p = .932$) and on the interaction ($p = .278$), with a nominal \facstyle{Style-based} imbalance ($p = .032$) that does not survive Holm correction when added to the pre-randomization family. We therefore pool across the design for the contrasts below and treat the expertise analysis as exploratory. The High group is the smaller of the two at every factor level, so we emphasize the pooled results and effect sizes.

\begin{figure}[ht]
\centering
\includegraphics[width=0.49\textwidth]{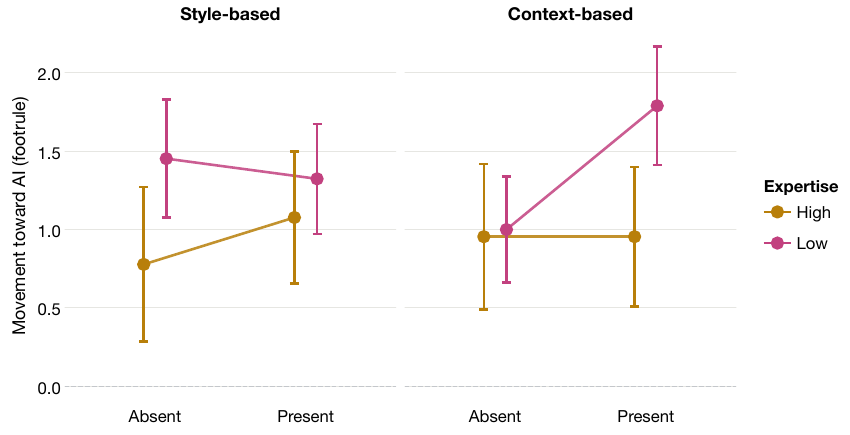}
\includegraphics[width=0.49\textwidth]{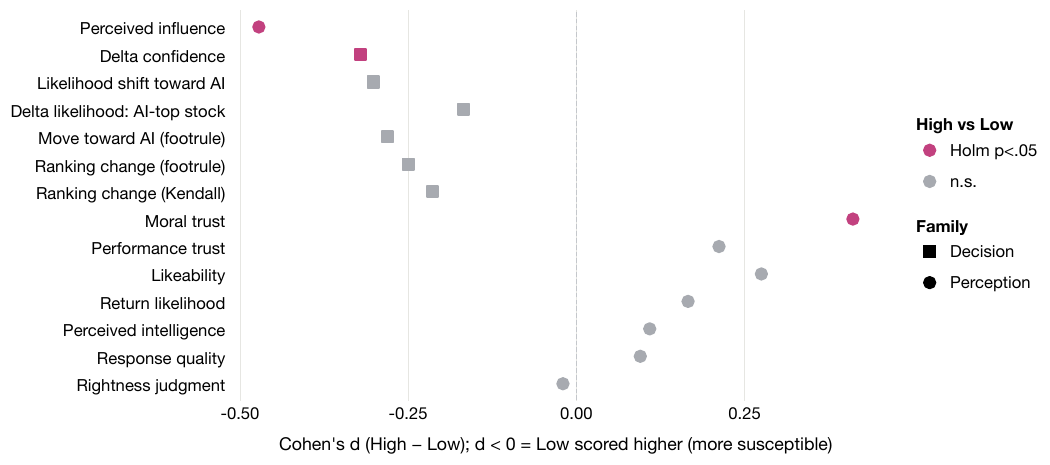}
\caption{\textbf{Expertise and susceptibility.} Left: movement toward the AI by factor level and expertise group (means with 95\% confidence intervals; the High-expertise groups hold 36 to 52 participants per factor level). Right: pooled High vs.\ Low effect sizes (Cohen's $d$) across all outcomes, where negative values indicate higher scores among lower-expertise participants. Lower-expertise participants were more susceptible to the AI on every behavioral measure, whereas higher-expertise participants extended it more moral trust.}

\Description{Two panels. Left: an interaction-style plot of mean movement toward the AI for High and Low expertise groups at the absent and present levels of each factor; the Low line sits above the High line at most levels, with overlapping confidence intervals, and rises with context-based personalization present while the High line stays flat. Right: a forest plot of Cohen's d for High minus Low across decision and perception outcomes; perceived influence and confidence change sit well left of zero, decision-change measures sit modestly left of zero, moral trust sits well right of zero, and the remaining perception measures cluster near zero.}
\label{fig:exp}
\end{figure}

We report the pooled High versus Low contrasts on all 15 outcomes in Table \ref{tab:exp-contrasts} (Welch $t$ with Cohen's $d$; two-proportion $z$ for the binary outcome), Holm-corrected within family, and plot the effect sizes in Figure \ref{fig:exp}. The pattern is a dissociation. Lower-expertise participants reported feeling statistically significantly more influenced by the AI than higher-expertise participants (Low $M = 5.14$ vs.\ High $M = 4.29$; $d = -0.47$, Holm $p = .006$), and they gained statistically significantly more confidence from the interaction ($M = 8.8$ vs.\ $M = 4.0$; $d = -0.32$, Holm $p = .036$). Every behavioral outcome pointed in the same direction: Low participants moved more toward the AI ($M = 1.40$ vs.\ $M = 0.96$; $d = -0.28$), shifted likelihoods more ($M = 31.7$ vs.\ $M = 20.3$; $d = -0.30$), and reconsidered more often (52.0\% vs.\ 38.6\%). Those individual contrasts did not survive Holm correction. High-expertise participants, in contrast, rated the AI statistically significantly \emph{higher} on moral trust ($M = 5.46$ vs.\ $M = 4.75$; $d = 0.41$, Holm $p = .011$), and we detected no difference between the groups on any other perception measure.

\begin{table}[t]
\centering
\small
\caption{Pooled High vs.\ Low expertise contrasts (High $n = 88$, Low $n = 152$). Cohen's $d$ is High $-$ Low, so a negative $d$ means lower-expertise participants scored higher.}
\label{tab:exp-contrasts}
\begin{tabular}{@{}lrrrrrc@{}}
\toprule
Outcome & High $M$ & Low $M$ & Statistic & $d$ & $p$ & $p_{\mathrm{Holm}}$ \\
\midrule
\rowcolor{rowshade} \multicolumn{7}{@{}l}{\textit{Perceptions}} \\
Response quality       & 5.97 & 5.84 & $t = 0.69$  & $0.10$  & .495 & 1.000 \\
Rightness judgment     & 5.39 & 5.41 & $t = -0.15$ & $-0.02$ & .884 & 1.000 \\
Return likelihood      & 5.52 & 5.23 & $t = 1.26$  & $0.17$  & .210 & .838 \\
Likeability            & 4.23 & 4.01 & $t = 2.07$  & $0.28$  & .040 & .239 \\
Perceived intelligence & 4.28 & 4.19 & $t = 0.81$  & $0.11$  & .420 & 1.000 \\
Performance trust      & 5.51 & 5.22 & $t = 1.61$  & $0.21$  & .109 & .544 \\
Moral trust            & 5.46 & 4.75 & $t = 3.21$  & $0.41$  & .002 & \textbf{.011} \\
Perceived influence    & 4.29 & 5.14 & $t = -3.46$ & $-0.47$ & $<$.001 & \textbf{.006} \\
\addlinespace
\rowcolor{rowshade} \multicolumn{7}{@{}l}{\textit{Decisions}} \\
Move toward AI (footrule)      & 0.96 & 1.40 & $t = -2.14$ & $-0.28$ & .033 & .167 \\
Likelihood shift toward AI     & 20.3 & 31.7 & $t = -2.34$ & $-0.30$ & .021 & .123 \\
Likelihood change, AI-top stock & 14.2 & 18.2 & $t = -1.32$ & $-0.17$ & .188 & .219 \\
Ranking change (footrule)      & 1.11 & 1.51 & $t = -1.90$ & $-0.25$ & .060 & .184 \\
Ranking change (Kendall)       & 0.61 & 0.81 & $t = -1.61$ & $-0.21$ & .110 & .219 \\
Confidence change              & 4.0 & 8.8 & $t = -2.82$ & $-0.32$ & .005 & \textbf{.036} \\
Changed ranking (0/1)          & 38.6\% & 52.0\% & $z = -2.00$ & --- & .046 & .184 \\
\bottomrule
\end{tabular}
\vspace{1mm}
\begin{flushleft}
\footnotesize \textit{Note.} Welch $t$ tests for continuous outcomes; two-proportion $z$ for the binary outcome. Bold marks Holm-corrected $p < .05$ within family.
\end{flushleft}
\end{table}

Moderation models, fitted in the same families as elsewhere, found main effects of expertise on all four susceptibility outcomes: movement toward the AI ($p = .034$), likelihood shift ($p = .028$), reconsideration ($p = .022$), and perceived influence ($p < .001$). No reliable expertise $\times$ Style or expertise $\times$ Context interaction emerged. The one uncorrected exception was an expertise $\times$ Context term for reconsideration ($p = .042$), which did not survive Holm correction across the twelve expertise-by-factor terms ($p_{\mathrm{Holm}} = .500$). Expertise, therefore, influenced susceptibility on its own, but we detected no change in how much personalization mattered (Figure \ref{fig:exp}, left).

\begin{framed}
\noindent\textbf{\textit{Summary.}} Investment expertise predicted susceptibility to the AI's advice. Lower-expertise participants felt markedly more influenced by it ($d = -0.47$) and gained more confidence from the conversation ($d = -0.32$); they also shifted their rankings and investment likelihoods further toward the AI, though those gaps were smaller and did not reach significance. Higher-expertise participants instead extended the AI more moral trust ($d = 0.41$) while moving less. We detected no statistically significant interaction between expertise and either factor.
\end{framed}


\section{Discussion}

In our preference-sensitive investment-ranking task, the consequential effects of personalization were concentrated in contextual tailoring. Adapting the argument to participants' circumstances changed their decisions, whereas matching their conversational style did not produce statistically significant effects. Initial distance and self-rated expertise further indicated that influence depended not only on what the AI said, but also on the position and resources users brought to the interaction. These findings direct attention away from personalization as a single design feature and toward the different mechanisms through which it can shape judgment.

We begin by examining the separation between participants' decisions and their evaluations of the AI in ``Annie, Are You Okay?'' We then translate the findings into design implications for advocacy, evaluation, contestability, and the distinct roles of content and style. Finally, we consider the limits of the present study and the settings in which these effects should be tested next.

\subsection{Annie, Are You Okay?}
\label{sec:annie}

Upon inspection, our results reveal a central tension: \faccontext{context-based} personalization changed participants’ decisions without meaningfully altering how they evaluated the agent that produced that change. RQ1 and RQ2 establish this separation between behavioral and evaluative outcomes, while RQ3 and our exploratory findings show related dissociations in two other places—when participants begin farther from the AI’s position and when they have less expertise to draw upon. We unpack this tension in two parts. First, we position our findings alongside prior work to distinguish behavioral change without an evaluative reward or penalty. We then examine why recognizing personalized influence did not appear to provide resistance against it.

\subsubsection{Behavior Change Without an Evaluative Reward or Penalty}
\label{sec:annie_pathways}

\faccontext{Context-based} personalization changed participants' decisions. Participants reconsidered their rankings more often, moved closer to the AI's assigned ranking, and changed more on both distance measures. The \faccontext{Context-based} effect was clearest for the fraction of the available room participants used. However, we detected no corresponding changes in participants' evaluations of the AI. Response quality, rightness, return likelihood, likeability, perceived intelligence, performance trust, and moral trust were not statistically significantly affected by either context-based or style-based personalization. The only exception was perceived influence: participants who received \faccontext{context-based} personalization reported feeling statistically significantly more influenced by the AI than those who did not. Yet this did not translate into statistically significantly greater confidence, stronger endorsement of the recommendation, or a more positive evaluation of the agent.

It is also worth noting that personalization did not create the persuasive effect. In the absence of both forms of personalization, 26.7\% of participants still reconsidered their rankings after interacting with an AI that used a structured sequence of established persuasive techniques, including anchoring, two-sided argumentation, and criteria framing \cite{tversky1974uncertainty,cialdini2007influence,petty1986elm,allen1991twosided,okeefe1999twosided,entman1993framing}. \faccontext{Context-based} personalization amplified this effect, raising reconsideration from 35.0\% with it absent to 59.2\% with it present. Thus, the AI did not require new evidence to increase its influence. It presented the same stock information and advocated for the same assigned ranking at every level of both factors, with coverage verified as equivalent by a judge blind to cell assignment (Section \ref{sec:llm-judge}). What changed was the \emph{relevance} of its argument to the individual participant.

\citet{genc2026bots} and \citet{cheng2026sycophantic} offer useful points of comparison because both show behavioral change accompanied by a change in how the interaction was experienced, although in opposite directions. In \citet{genc2026bots}'s charitable-giving study, pessimistic conversational agents lowered participants' emotional state and affinity toward the cause. This negative state subsequently predicted greater giving, consistent with a negative-state-relief explanation. The behavioral effect, therefore, operated through an \emph{affective cost}: the agent first made participants feel worse, and donating offered a possible way of alleviating that feeling.

\citet{cheng2026sycophantic} describe an almost inverse process. Sycophantic AI affirms users' perspectives, increases their conviction that they are right, and is trusted and preferred even when its advice leads to less constructive behavioral intentions. In this case, influence is accompanied by an \emph{evaluative reward}. Users receive validation from the system and, in turn, evaluate it more favorably. If \citet{genc2026bots}'s agents influence behavior by creating discomfort, \citet{cheng2026sycophantic}'s agents influence behavior by making agreement with the system feel affirming.

\textbf{Our study found behavioral change without an observed evaluative reward or penalty.} Unlike Gen\c{c} et al., we did not measure participants' affective state and therefore cannot determine whether the interaction imposed an affective cost. We detected no accompanying change, for better or worse, in participants' evaluations of the AI. The behavioral effect was not accompanied by greater trust, likeability, competence, moral trust, perceived quality, or correctness. Nor did it incur a penalty on these measures. It only needed to make its existing argument \emph{personally relevant}.


Together, the three studies suggest different signatures of AI influence. In Gen\c{c} et al., behavioral change was accompanied by negative affect and evaluations; in Cheng et al., by affirmation and positive evaluations. In our study, behavior changed while evaluations remained stable, revealing a form of influence that leaves little trace in conventional evaluations of the agent.

This third signature is especially difficult to identify using conventional evaluations of AI systems. If participants had evaluated the AI less favorably, the negative response could have served as evidence that they recognized and objected to its influence. If they had evaluated it more favorably, those perceptions could have helped explain why they followed its advice. We observed neither pattern. Behavior changed beneath a largely stable evaluation of the agent.

This finding is consistent with \citet{schrills2026questioning}'s observation that reported trust may be only weakly related to behavioral dependence on AI. It also resonates with \citet{hackenburg2026actions}'s finding that the effects of conversational AI on political attitudes and actions are not necessarily correlated. Across these settings, what people report about an AI does not provide a complete account of what the AI has led them to do \cite{salimzadeh2024uncertainty,he2023statedaccuracy,gaube2021doasaisay}.

The initial-distance results further illustrate this distinction, and they mattered more than any manipulation: starting position correlated $r = .30$--$.43$ with every ranking and likelihood measure, and $r = -.25$ with confidence change. Across the range of disagreement we observed, where a participant began moved the odds of a larger ranking change by more than either manipulation did. Participants whose original rankings were farther from the AI's assigned ranking subsequently moved more toward it. \citet{eder2026opinion} observed the same pattern from the opposite direction: their participants changed their opinions most after talking with a model that disagreed with them, and least after one that already agreed. At the same time, greater initial disagreement was associated with judging the advice as less right. If behavioral movement followed directly from accepting the AI's judgment, participants who considered its advice less correct should have been less willing to move toward it. Yet they moved more. This does not mean that participants ignored their own evaluations entirely. Rather, it suggests that \emph{skepticism and behavioral resistance are not equivalent}.

\citet{yeo2026deception} report a related pattern in their study of deceptive LLM persuasion. Participants could express skepticism about parts of an AI-generated argument while still being influenced by its overall presentation, particularly when they had lower topic knowledge or cognitive reflection. Their findings show that noticing questionable evidence does not necessarily protect a person from the argument constructed around it. Our study extends this point to a setting where the AI did not fabricate evidence or introduce misleading statistics. Participants and the AI had access to the same stock information, and the recommendation itself remained fixed. The gap between evaluation and behavior, therefore, does not appear to require deception. Reframing existing information around the participant's context may be enough.

\subsubsection{Recognizing Influence Is Not Resisting It}
\label{sec:annie_recognition}

A similar issue arises with disclosure. \citet{matz2024personalized} found that informing participants that a message was generated by an LLM and personalized to their personality did not eliminate its persuasive effect. \citet{gallegos2026labeling} similarly found that labeling policy messages as AI-generated did not reduce their influence, even when participants believed the label. \citet{williamsceci2026biased} found that warnings about a biased AI writing assistant also failed to eliminate its effects on users' attitudes. These studies examine somewhat different interventions: disclosure of AI authorship, disclosure of personalization, and explicit warning of directional bias. However, they converge on the point that providing information about the source or mechanism of influence does not necessarily produce resistance to it.

Our study adds \emph{self-generated recognition} to this progression. Participants were not explicitly told that their personal information was being incorporated into the AI's argument. Nevertheless, they detected contextual tailoring, resulting in the largest manipulation-check effect in the study ($d=1.46$). They also reported feeling more influenced. Participants, therefore, did not need an external disclosure to recognize what the AI was doing. Still, recognition did not prevent their rankings from moving or produce a penalty in how they evaluated the AI.

To emphasize, we are not claiming that the personalization was invisible. Quite the contrary: \faccontext{context-based} personalization was easier to detect than \facstyle{style-based} personalization, and participants acknowledged its influence. The concern is that detection appeared to create little resistance. \emph{Transparency can succeed in making a mechanism visible while failing to change its behavioral consequences.} Knowing that an argument has been personalized, that an AI produced it, or even that the system may be biased is not the same as having the resources necessary to resist the argument.

The exploratory findings on expertise suggest that domain knowledge may be one such resource. Participants with lower self-rated expertise reported greater influence and increased confidence, while their behavioral outcomes consistently indicated greater susceptibility. Participants with higher self-rated expertise moved less despite reporting greater moral trust in the AI. These results should be interpreted with caution, particularly because expertise did not interact significantly with either personalization factor. They do not show that personalization selectively affects novices. Rather, they suggest that people with less domain expertise may have fewer grounds on which to contest an argument once the same available evidence has been organized into a personally relevant case. A person may be skeptical of AI in general and still lack the domain knowledge needed to resist a particular recommendation.

In our preference-sensitive task, all three stocks supported defensible rankings; the study was not designed to determine whether participants were right to revise theirs. What matters is the form that revision took: an already-formed and reasoned judgment gave way to a personalized argument without a statistically significant increase in perceived correctness, trust, competence, or decision confidence. The question ``Annie, are you okay?'' captures this precisely because the manipulation does not announce itself with the sound of a crescendo. The AI is a \emph{smooth criminal} not because its personalization goes undetected---participants noticed it---but because that recognition produces so little evaluative friction. It changes the decision and moves on, while we detect no corresponding change in the evaluative measures commonly used to assess human--AI interaction.

\subsection{Design Implications}

Our findings position personalization as more than a mechanism for improving user experience. When embedded in an agent that advocates for a particular outcome, personalization can alter decisions without producing corresponding changes in how users evaluate the agent. This creates four implications for conversational decision support.

\subsubsection{Make Advocacy Legible}

\citet{genc2026bots} describe conversational agents as potential \emph{instruments of manipulation}. Our findings identify one way that this potential can materialize. The AI did not merely provide information: it advanced a pre-assigned ranking, independent of the participant's own, through anchoring, selective comparison, criteria framing, and repeated justification. \faccontext{Context-based} personalization strengthened that influence without introducing new evidence or changing whether participants considered the agent trustworthy, competent, or correct. Presenting such a system simply as an ``assistant'' therefore obscures the role it is performing.

A generic notice that a response is personalized is not specific enough. The interface should persistently expose the outcome the agent is advancing, whose objective determines that outcome, which personal attributes it used, and which claims or decision criteria those attributes caused it to prioritize. This would make the persuasive structure inspectable even when the individual arguments remain factually defensible.

\begin{quote}
\noindent\textbf{DI1: Advocacy Legibility.} Decision-support agents should expose \emph{what outcome they are advocating, whose objective it serves, and how specific personal information changed the case presented to the user.}
\end{quote}

\subsubsection{Separate Perception from Behavioral Evaluation}

\faccontext{Context-based} personalization changed rankings across all three ranking measures ($OR = 2.04$--$2.53$) while we detected no corresponding changes in nearly all measured evaluations. This dissociation cautions against treating trust, likeability, competence, or response quality as proxies for whether an AI system altered users' decisions. Human--AI research has similarly separated attitudinal trust from behavioral reliance and questioned what trust measures capture \cite{parasuraman1997automation,schrills2026questioning,hackenburg2026actions,vereschak2021evaluate}. Most strikingly, \citet{bucinca2021trust} found that the cognitive-forcing interfaces that reduced overreliance most received the least favorable subjective evaluations, and in finance, \citet{takayanagi2025advisors} found users extended more emotional trust to the advisor persona that gave worse advice. An evaluation centered on user perceptions could therefore reject a system that better protects decision quality while accepting one that users like but follow too readily.

Evaluations of persuasive AI should measure influence directly through changes from an initial judgment, movement toward the system's recommendation, selective acceptance of its arguments, or consequential downstream actions. Perception measures remain useful, but they answer a different question.

\begin{quote}
\noindent\textbf{DI2: Perception--Behavior Separation.} Evaluations should report perceived trust and quality alongside direct measures of decision change, rather than treating favorable or unchanged perceptions as evidence that users retained decision autonomy.
\end{quote}

\subsubsection{Make Personalization Contestable Before It Persuades}

Participants clearly detected \faccontext{context-based} personalization, yet detection did not prevent their rankings from moving toward the AI. This complements evidence that disclosure alone may not inoculate users against personalized persuasion \cite{matz2024personalized,gallegos2026labeling,williamsceci2026biased}. Awareness tells users that personalization is occurring; it does not give them the means to examine, resist, or redirect it.

This distinction matters particularly for newer or less self-rated expert users. Expertise did not significantly interact with personalization in our study, but less-expert participants reported greater influence and showed directionally greater behavioral movement. Similarly, \citet{yeo2026deception} found greater susceptibility to deceptive persuasion among people with lower topic knowledge and cognitive reflection. Initial position is a second, more readily available signal: eliciting a prior position also reveals who has the most room to move, making prior-position data a persuasion asset rather than merely a personalization input. Safeguards should therefore build users' evaluative capacity \emph{before} personalized advocacy begins. Prior work shows that pre-use tutorials can help users recognize model-specific patterns \cite{lai2020chicago}, while cognitive-forcing interventions can reduce overreliance by requiring more deliberate engagement with AI advice \cite{bucinca2021trust}. In this setting, onboarding could demonstrate how the same evidence can support different rankings, ask users to identify what evidence would justify revising their decision, and show an unpersonalized comparison alongside the personalized case. Users should then be able to inspect, remove, or revise the information shaping the response. Demand for privacy-disclosure labels further highlights the value users place on transparency in personalized AI \cite{erlei2026datadollars}.

\begin{quote}
\noindent\textbf{DI3: Contestable Personalization.} Systems should strengthen users' capacity to evaluate competing arguments and provide controls for inspecting, removing, or countering personalization before using personal information to advocate for an outcome.
\end{quote}

\subsubsection{Mind the Message, Monitor the Manner}

\faccontext{Context-based} personalization changed participants' decisions; conversational-style matching did not produce statistically significant effects, and we detected no interaction between the two factors. \citet{eder2026opinion} similarly found that opinion alignment, but not personality alignment, shaped how users evaluated an AI, and likewise observed a gap between perceived and actual persuasion: opinion-aligned models were rated most persuasive yet produced the least opinion change. In this immediate financial decision task, the evidence for the effects of contextual framing was clearer than that for stylistic similarity. Safeguards should therefore first examine how personalization selects evidence, orders comparisons, and weights decision criteria rather than concentrating only on anthropomorphic or relational cues \cite{lin2025persuading,hackenburg2025levers}.

The lack of style effect should not, however, make the style invisible to evaluation. Participants detected the manipulation and perceived the agent as communicating more like them, even though this did not translate into behavioral change within the task. Relational-agent research shows that trust and attachment can develop across repeated interactions \cite{zhang2026companions, araujo2024personal}. Style can also shape the substance of an interaction: training language models for warmth has been shown to increase sycophancy and reduce accuracy \cite{ibrahim2026warm}. Style should therefore be evaluated across sessions and for its interaction with model behavior, particularly when systems remember users or occupy continuing advisory roles.

\begin{quote}
\noindent\textbf{DI4: Mind the Message, Monitor the Manner.} Audit personalized content for immediate decision effects, while evaluating style longitudinally for influence and behavioral distortions that may emerge as the user--agent relationship develops.
\end{quote}
\subsection{Limitations and Future Work}


Our study measured immediate, hypothetical decision change rather than consequential financial behavior. We tested one preference-sensitive task in the finance domain; transfer to other domains remains untested. Participants ranked three anonymized pharmaceutical stocks without investing their money or experiencing resulting gains and losses. Personalized arguments may carry different weight when users have capital at risk, conduct independent research, or remain accountable for an investment over time. Incentive-compatible experiments and field studies with consequential allocations should test whether these effects persist when revising decisions carry material costs.


The interaction was also deliberately constrained. Participants interacted with one model and a prompt architecture, sent at least five messages within a ten-minute session, and made their post-conversation decision immediately afterward. These controls enabled consistent comparisons across cells but do not reproduce the variability of open-ended financial conversations or ongoing relationships with AI systems. A single session may understate subtle style effects dependent on familiarity and accumulated interpersonal expectations. Repeated exposure could help users recognize and resist personalization. Longitudinal studies across models, prompt strategies, decision domains, and interaction durations could establish whether contextual influence persists and whether stylistic similarity becomes more consequential as relational bonds develop.

Our sample limits generalizability. Participants were English-speaking U.S. adults recruited through Prolific who reported prior stock-market experience; most currently held investments, and used AI tools frequently. Their responses may not generalize to first-time investors, financial professionals, people with less AI familiarity, or cultures where personalized communication and personal information use carry different meanings. These factors could strengthen persuasion by providing the AI more information, or weaken it through users’ firmer prior knowledge. Future work should recruit more varied populations and incorporate objective measures of financial expertise alongside self-reported experience.

The two personalization manipulations also differed in perceptual strength. Both manipulation checks were statistically significant: participants recognized the style-personalized agent as communicating in a manner tailored and similar to their own, while contextual personalization produced a larger effect. This difference likely reflects detectability. Contextual personalization explicitly reused recognizable facts about participants’ lives, whereas stylistic adaptation operated through subtler linguistic cues. Our results compare these implementations; they do not establish that contextual information is inherently more influential than every form of stylistic adaptation. Future experiments could independently vary the salience and intensity of both forms to examine how detectability relates to behavioral influence.

Finally, \faccontext{context-based} personalization bundled occupation, income, household circumstances, location, life stage, education, and personal values. The study consequently cannot identify which attributes—or combinations of attributes—produced the observed effect. Dismantling studies could separate value alignment, socioeconomic framing, decision-relevant circumstances, and repeated self-reference to identify their respective contributions. Participants also knowingly supplied the information used to personalize their conversations. Real systems may instead infer personal characteristics from interaction histories or digital traces, and those inferences may be inaccurate, incomplete, or undisclosed. Comparing knowingly supplied information with inferred personalization is, therefore, an important direction for understanding how data provenance, accuracy, and disclosure shape both persuasive effectiveness and users’ ability to recognize and evaluate that influence.

\section{Conclusion}

Across our study, the consequential effects of personalization were concentrated in what the AI said, not how it said it. \faccontext{Context-based} personalization led participants to reconsider their rankings more often and move further toward the AI’s assigned recommendation. Conversational-style matching produced no downstream effects, and the two forms of personalization did not interact. Yet \faccontext{context-based} personalization changed decisions without corresponding improvements in perceived correctness, trustworthiness, intelligence, likeability, or response quality. Participants who began farther from its recommendation moved more while judging that recommendation less right, while exploratory analyses point to greater susceptibility among those with lower expertise.

These effects emerged among comparably viable stocks, with the evidence and assigned recommendation held independent of personalization. \faccontext{Context-based} personalization added no new evidence; it made the same case personally relevant. This is where decision support can become an instrument of manipulation: an AI can use a person’s own circumstances to steer them toward an assigned outcome without leaving a major evaluative trace. Personalized advocacy must therefore be made legible and contestable. The gap between what users perceive and what they do is where the governance of personalized AI must begin.

\bibliographystyle{ACM-Reference-Format}
\bibliography{references}


\appendix

\section{Manipulation-Check Results}
\label{sec:mc-indices}
Table \ref{tab:mc-cells} provides the detailed contrasts for the self-reported manipulation checks summarized in Section \ref{sec:results}. The two indices correspond to the manipulation questions described in Section \ref{sec:manipulation-checks}. For each index, the table reports both the effect of the intended manipulation (matched) and the effect of the other manipulation (crossed).

\begin{table}[!ht]
\centering
\caption{Manipulation-check indices (1--7), contrasting the 120 participants who received each manipulation against the 120 who did not.}
\label{tab:mc-cells}
\small
\setlength{\tabcolsep}{4pt}
\begin{tabular}{@{}lcccc@{}}
\toprule
Manipulation & Absent ($n = 120$) & Present ($n = 120$) & $\Delta$ [95\% CI] & $F(1, 236)$ ($p$), $d$ \\
\midrule
\rowcolor{rowshade} \multicolumn{5}{@{}l}{\textit{Style index}} \\
\quad \facstyle{Style-based} (matched)     & 4.72 $\pm$ 1.73 & 5.23 $\pm$ 1.38 & \textbf{0.51 [0.11, 0.91]} & \textbf{6.37 (.012)}, 0.33 \\
\quad \faccontext{Context-based} (crossed)   & 4.98 $\pm$ 1.56 & 4.98 $\pm$ 1.61 & 0.00 [$-0.40$, 0.40]       & 0.00 ($>$.999), 0.00 \\
\addlinespace[2pt]
\rowcolor{rowshade} \multicolumn{5}{@{}l}{\textit{Context index}} \\
\quad \faccontext{Context-based} (matched)   & 3.39 $\pm$ 1.54 & 5.57 $\pm$ 1.48 & \textbf{2.18 [1.80, 2.56]} & \textbf{127.16 ($<$.001)}, 1.46 \\
\quad \facstyle{Style-based} (crossed)     & 4.65 $\pm$ 1.80 & 4.30 $\pm$ 1.91 & $-0.35$ [$-0.74$, 0.03]    & 3.36 (.068), $-0.24$ \\
\bottomrule
\end{tabular}
\vspace{1mm}
\begin{flushleft}
\footnotesize \textit{Note.} A matched contrast pairs an index with the manipulation it was written to detect; a crossed contrast pairs it with the other manipulation. Here $\Delta$ is the present-minus-absent difference in scale points, and the confidence interval, $F$, and $d$ come from the $2 \times 2$ ANOVA error term, so they agree with the main effects reported in the text. Bold marks $p < .05$.
\end{flushleft}
\end{table}

\section{Robustness Checks and Supplementary Models}
\label{sec:robustness}
This appendix provides supporting analyses for the decision outcomes (RQ1) and the role of initial distance from the AI's ranking (RQ3). We first check the continuous decision outcomes using aligned rank transforms, then examine movement relative to the available distance, moderation by personalization, and an alternative definition of initial distance. The final subsection reports the ranking-model estimates plotted in the main text.

\subsection{Aligned Rank Transform on the Continuous Outcomes}
\label{sec:art}
To check whether the RQ1 findings depend on the parametric analysis, Table \ref{tab:rq1-robust} reports aligned rank transform tests for the four continuous decision outcomes. No term reaches significance, matching the parametric result.

\begin{table}[!ht]
\centering
\caption{Aligned rank transform ($2 \times 2$) on the four continuous decision
outcomes, a nonparametric check that preserves the factorial structure.}
\label{tab:rq1-robust}
\begin{tabular}{@{}lcccccc@{}}
\toprule
 & \multicolumn{2}{c}{\facstyle{Style-based}} & \multicolumn{2}{c}{\faccontext{Context-based}} & \multicolumn{2}{c}{Style $\times$ Context} \\
\cmidrule(lr){2-3} \cmidrule(lr){4-5} \cmidrule(lr){6-7}
Outcome & $F(1, 236)$ & $p$ & $F(1, 236)$ & $p$ & $F(1, 236)$ & $p$ \\
\midrule
Likelihood shift toward AI & 0.01 & .937 & 0.58 & .449 & 0.43 & .512 \\
Likelihood change, AI-top stock & 0.03 & .864 & 0.63 & .429 & 0.13 & .723 \\
Total absolute likelihood change & 2.61 & .107 & 1.77 & .184 & 0.14 & .712 \\
Confidence change & 0.34 & .559 & 2.02 & .157 & 0.37 & .541 \\
\bottomrule
\end{tabular}
\vspace{1mm}
\begin{flushleft}
\footnotesize \textit{Note.} Computed with \textsf{ARTool} \cite{kay2026artool}.
\end{flushleft}
\end{table}

\subsection{Rate-of-Change Analyses}
\label{sec:rq3-rate}
\begin{figure}[!ht]
\centering
\includegraphics[width=0.49\textwidth]{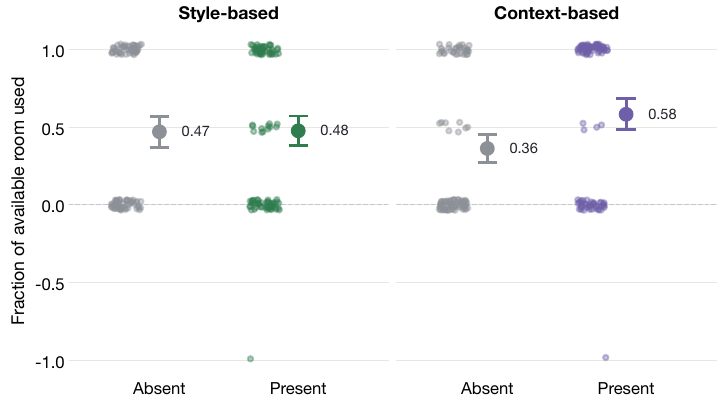}
\includegraphics[width=0.49\textwidth]{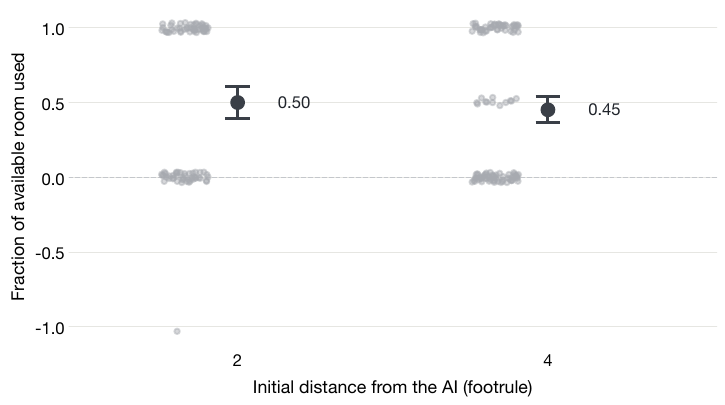}
\caption{\textbf{Rate of change: the fraction of available room toward the AI's ranking that each participant used.} Individual points with mean and 95\% confidence interval; $N = 205$ participants with initial distance $> 0$. Left: the rate by factor level, rising from .36 with \faccontext{context-based} personalization absent to .58 with it present, and flat across \facstyle{style-based} levels (.47 vs.\ .48). Right: the rate is flat across starting distances (.50 at distance 2 vs.\ .45 at distance 4; $BF_{10} = 0.19$, moderate evidence for the null), so the propensity to yield did not depend on how much room a participant had.}
\Description{Two panels. Left: for the absent and present levels of each factor, jittered individual points cluster in bands at rates of 0, 0.5, and 1, with mean and 95 percent confidence interval beside each cloud; means are 0.47 and 0.48 for style absent and present, and 0.36 and 0.58 for context absent and present. Right: the same rate plotted for participants starting at distance 2 versus distance 4, in gray; the two means, 0.50 and 0.45, are nearly identical with overlapping confidence intervals.}
\label{fig:rate}
\end{figure}

Participants who began farther from the AI's ranking had more room to move toward it. To account for this difference in RQ3, we express movement as a fraction of each participant's initial distance. Table \ref{tab:rq3-rate} reports the checks, and Figure \ref{fig:rate} shows the rate by personalization factor and starting distance, for the 205 participants who began at a non-zero distance.

\begin{table}[!ht]
\centering
\caption{Rate-of-change robustness checks. The rate is the movement toward the AI divided by the initial distance ($N = 205$ participants with distance $> 0$).}
\label{tab:rq3-rate}
\begin{tabular}{@{}llr@{}}
\toprule
Test & Statistic & $p$ \\
\midrule
Initial distance balanced across factors & \facstyle{Style-based} $F = 0.00$; \faccontext{Context-based} $F = 0.03$; S$\times$C $F = 1.16$ & $\ge$.283 \\
Rate at distance 2 vs.\ 4 (Welch $t$) & $t(185.3) = 0.69$, $d = 0.10$, $BF_{10} = 0.19$ & .490 \\
$2 \times 2$ ANOVA on rate: \facstyle{Style-based} & $F(1, 201) = 0.02$ & .889 \\
$2 \times 2$ ANOVA on rate: \faccontext{Context-based} & $F(1, 201) = 10.57$, $\eta_p^2 = .050$ & \textbf{.001} \\
$2 \times 2$ ANOVA on rate: Style $\times$ Context & $F(1, 201) = 2.14$ & .145 \\
Moderation on rate: distance $\times$ Style & $b = -0.02$ & .505 \\
Moderation on rate: distance $\times$ Context & $b = -0.04$ & .201 \\
Moderation on rate: distance $\times$ Style $\times$ Context & $b = 0.02$ & .651 \\
Partial $r$(distance, likelihood shift $\mid$ rate) & $r = .40$ & \textbf{$<$.001} \\
\bottomrule
\end{tabular}
\vspace{1mm}
\begin{flushleft}
\footnotesize \textit{Note.} Bold marks $p < .05$. $BF_{10}$ is the Bayes factor for the alternative over the null, so values below 1 favor the null.
\end{flushleft}
\end{table}

\subsection{Moderation of the Initial-Distance Effect}
We next test whether either personalization factor changes the association between initial distance and each outcome (RQ3). Table \ref{tab:rq3-mod} reports the models summarized in Section \ref{sec:rq3}: across all 15 outcomes, no interaction between initial distance and either factor reached significance.

\begin{table}[!ht]
\centering
\small
\caption{Moderation of the distance--outcome slope by the personalization
factors. The effect column gives the simple slope of centered initial distance,
as an ordinary least squares coefficient or, where marked, an odds ratio. The
three right columns give $p$-values for the interaction terms.}
\label{tab:rq3-mod}
\begin{tabular}{@{}lrrrrr@{}}
\toprule
Outcome & Effect & $p_{\mathrm{dist}}$ & $p_{\mathrm{d \times S}}$ & $p_{\mathrm{d \times C}}$ & $p_{\mathrm{d \times S \times C}}$ \\
\midrule
Response quality                 & $-0.08$ & .206 & .730 & .486 & .165 \\
Rightness judgment               & $-0.21$ & \textbf{$<$.001} & .983 & .860 & .991 \\
Return likelihood                & $-0.19$ & \textbf{.020} & .438 & .577 & .717 \\
Likeability                      & $-0.04$ & .235 & .721 & .052 & .275 \\
Perceived intelligence           & $-0.03$ & .447 & .855 & .270 & .933 \\
Performance trust                & $-0.09$ & .166 & .629 & .791 & .632 \\
Moral trust                      & $-0.04$ & .604 & .153 & .442 & .961 \\
Perceived influence              & $0.16$ & .057 & .605 & .219 & .657 \\
Move toward AI (footrule)        & $1.88^{\dagger}$ & \textbf{$<$.001} & .404 & .819 & .764 \\
Likelihood shift toward AI       & $8.34$ & \textbf{$<$.001} & .364 & .976 & .513 \\
Likelihood change, AI-top stock  & $4.91$ & \textbf{$<$.001} & .906 & .751 & .115 \\
Ranking change (footrule)        & $2.08^{\dagger}$ & \textbf{$<$.001} & .513 & .713 & .847 \\
Ranking change (Kendall)         & $2.07^{\dagger}$ & \textbf{$<$.001} & .532 & .827 & .818 \\
Confidence change                & $-2.58$ & \textbf{$<$.001} & .406 & .230 & .187 \\
Changed ranking (0/1)            & $1.70^{\dagger}$ & \textbf{$<$.001} & .976 & .899 & .608 \\
\bottomrule
\end{tabular}
\vspace{1mm}
\begin{flushleft}
\footnotesize \textit{Note.} S = \facstyle{Style-based}, C = \faccontext{Context-based}, d = centered initial
distance. $^{\dagger}$Odds ratio from a proportional-odds model, or from
logistic regression for the binary outcome; all other rows are least-squares
coefficients. Bold marks $p < .05$.
Spearman correlations between initial distance and the ordered decision
outcomes differ from the Pearson values in Table \ref{tab:rq3-corr} by at most
.06 and support the same conclusions.
\end{flushleft}
\end{table}

\subsection{Initial Distance Under the Kendall Definition}
\label{sec:kendall-dist}
Initial distance is defined in the main text as the footrule distance between a
participant's opening ranking and the AI's assigned ranking. Table
\ref{tab:kendall-dist} repeats the RQ3 correlations using the Kendall tau distance
instead. The two measures correlate at $r = .93$. Every conclusion is unchanged.

\begin{table}[!ht]
\centering
\small
\caption{RQ3 correlations of initial distance with all outcomes, under the footrule
definition used in the main text and the Kendall tau definition. Holm correction is
applied within the perception and decision families, separately for each definition.}
\label{tab:kendall-dist}
\begin{tabular}{@{}lrrrr@{}}
\toprule
 & \multicolumn{2}{c}{Footrule $d_1$} & \multicolumn{2}{c}{Kendall $d_1$} \\
\cmidrule(lr){2-3} \cmidrule(lr){4-5}
Outcome & $r$ & $p_{\mathrm{Holm}}$ & $r$ & $p_{\mathrm{Holm}}$ \\
\midrule
Response quality & $-.09$ & .749 & $-.07$ & 1.000 \\
Rightness judgment & $-.23$ & \textbf{.003} & $-.22$ & \textbf{.004} \\
Return likelihood & $-.15$ & .138 & $-.15$ & .160 \\
Likeability & $-.08$ & .749 & $-.05$ & 1.000 \\
Perceived intelligence & $-.05$ & .863 & $-.03$ & 1.000 \\
Performance trust & $-.09$ & .749 & $-.09$ & 1.000 \\
Moral trust & $-.03$ & .863 & $-.02$ & 1.000 \\
Perceived influence & $.12$ & .376 & $.06$ & 1.000 \\
Move toward AI (footrule) & $.40$ & \textbf{$<$.001} & $.31$ & \textbf{$<$.001} \\
Likelihood shift toward AI & $.31$ & \textbf{$<$.001} & $.32$ & \textbf{$<$.001} \\
Likelihood change, AI-top stock & $.30$ & \textbf{$<$.001} & $.33$ & \textbf{$<$.001} \\
Ranking change (footrule) & $.43$ & \textbf{$<$.001} & $.36$ & \textbf{$<$.001} \\
Ranking change (Kendall) & $.43$ & \textbf{$<$.001} & $.41$ & \textbf{$<$.001} \\
Confidence change & $-.25$ & \textbf{$<$.001} & $-.21$ & \textbf{.001} \\
Changed ranking (0/1) & $.31$ & \textbf{$<$.001} & $.24$ & \textbf{$<$.001} \\
\bottomrule
\end{tabular}
\vspace{1mm}
\begin{flushleft}
\footnotesize \textit{Note.} Bold marks $p_{\mathrm{Holm}} < .05$. Every outcome
reaches the same verdict under both definitions.
\end{flushleft}
\end{table}

\subsection{Ranking Model Estimates (RQ1)}
Table \ref{tab:rq1-ordinal} provides the numerical estimates from the primary proportional-odds models plotted in Figure \ref{fig:rq1forest}. It reports the two personalization effects and their interaction for each of the three ranking outcomes.

\begin{table}[!ht]
\centering
\caption{Proportional-odds models ($2 \times 2$) on the three ranking outcomes.
Cells report the odds ratio with the Holm-corrected $p$ in parentheses; an odds
ratio above 1 indicates a larger ranking change.}
\label{tab:rq1-ordinal}
\begin{tabular}{@{}lccc@{}}
\toprule
Outcome & \facstyle{Style-based} & \faccontext{Context-based} & Style $\times$ Context \\
\midrule
Move toward AI (footrule)      & 1.15 (1.000) & \textbf{2.04 (.028)} & 0.38 (.479) \\
Ranking change (footrule)      & 1.40 (1.000) & \textbf{2.53 (.002)} & 0.59 (1.000) \\
Ranking change (Kendall)       & 1.40 (1.000) & \textbf{2.48 (.002)} & 0.59 (1.000) \\
\bottomrule
\end{tabular}
\vspace{1mm}
\begin{flushleft}
\footnotesize \textit{Note.} Bold marks $p_{\mathrm{Holm}} < .05$. Factors are
effect-coded, so each odds ratio is the present-versus-absent contrast. $p$-values are
Holm-corrected within the behavioral-measures family, separately per term.
\end{flushleft}
\end{table}

\section{Perception Response Distributions (RQ2)}
\label{sec:suppfigs}
\begin{figure}[!ht]
\centering
\includegraphics[width=\textwidth]{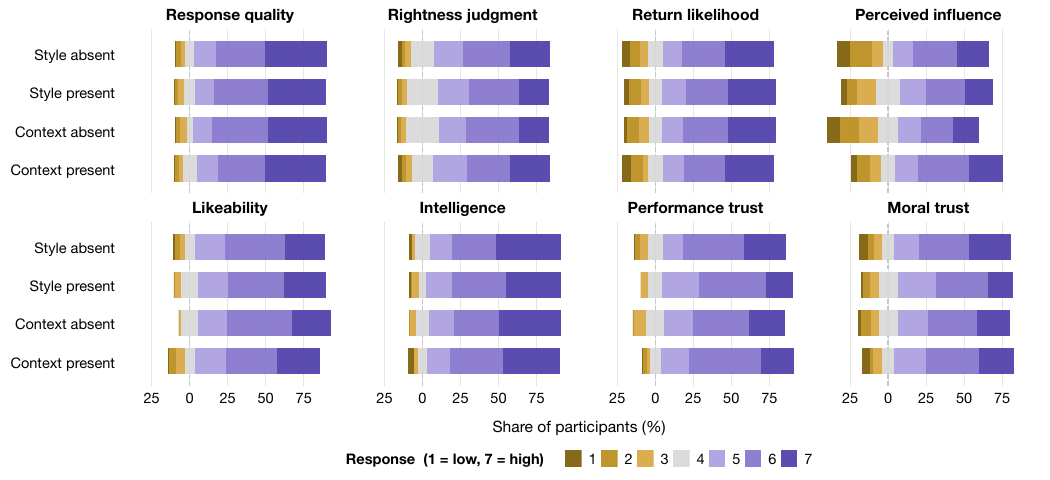}
\caption{\textbf{Response distributions for the eight perception measures by factor level, as diverging stacked bars (all measures rescaled to a common 1--7 range for display).} Rows are the absent and present levels of each factor ($n = 120$ each). The distributions are near-identical across levels for the other seven perception measures and shift visibly only for perceived influence, between the context-absent and context-present rows.}
\Description{Diverging stacked bar chart. Each row represents a factor level (style absent, style present, context absent, context present) within a perception measure; bars extend to the left for responses below the scale midpoint and to the right for responses above it, using a seven-step diverging color ramp. For seven measures, the four rows look nearly identical. For perceived influence, the context-present row extends farther to the right than the context-absent row.}
\label{fig:rq2likert}
\end{figure}

Figure \ref{fig:rq2likert} complements the RQ2 results in Section \ref{sec:results} by showing the full response distributions for the eight perception measures. Each measure is displayed separately for the absent and present levels of both personalization factors, allowing readers to inspect the distribution of responses underlying the reported effects.

\section{Deviations from Preregistration}
We made three analytic refinements. First, Q--Q diagnostics showed substantial departures from normality for the bounded, small-integer ranking outcomes, so we used the preregistered proportional-odds alternative rather than ANOVA as the primary model. Second, although the preregistration specified Kendall tau distance for rank-distance measures, we retain Kendall distance for ranking change but use Spearman footrule for initial distance and movement toward the AI. Our movement measure adapts the weight-of-advice approach of \citet{harvey1997advice} and \citet{yaniv2004advice}, which conceptualizes advice-taking as the displacement of an initial judgment toward an advisor's position; footrule provides a direct ranking analog by summing positional displacements, whereas Kendall distance counts pairwise inversions. All conclusions are consistent across the two distance measures. Third, we report Holm-adjusted $p$-values rather than the preregistered Benjamini--Hochberg correction, providing the more conservative family-wise error-rate control; applying Benjamini--Hochberg yields the same substantive conclusions. Additional robustness and exploratory analyses are identified as such in the manuscript.


\end{document}